# Geochemistry constrains global hydrology on Early Mars

Edwin S. Kite[1,*] and Mohit Melwani Daswani[1,2]

1. University of Chicago, Chicago, IL.
2. Now at: Jet Propulsion Laboratory, Caltech, Pasadena, CA.
*   kite@uchicago.edu

**Abstract.**

Ancient hydrology is recorded by sedimentary rocks on Mars. The most voluminous sedimentary rocks that formed during Mars' Hesperian period are sulfate-rich rocks, explored by the *Opportunity* rover from 2004-2012 and soon to be investigated by the *Curiosity* rover at Gale crater. A leading hypothesis for the origin of these sulfates is that the cations were derived from evaporation of deep-sourced groundwater, as part of a global circulation of groundwater. Global groundwater circulation would imply sustained warm Earthlike conditions on Early Mars. Global circulation of groundwater including infiltration of water initially in equilibrium with Mars' $CO_2$ atmosphere implies subsurface formation of carbonate. We find that the $CO_2$ sequestration implied by the global groundwater hypothesis for the origin of sulfate-rich rocks on Mars is 30-5000 bars if the *Opportunity* data are representative of Hesperian sulfate-rich rocks, which is so large that (even accounting for volcanic outgassing) it would bury the atmosphere. This disfavors the hypothesis that the cations for Mars' Hesperian sulfates were derived from upwelling of deep-sourced groundwater. If, instead, Hesperian sulfate-rich rocks are approximated as pure Mg-sulfate (no Fe), then the $CO_2$ sequestration is 0.3-400 bars. The low end of this range is consistent with the hypothesis that the cations for Mars' Hesperian sulfates were derived from upwelling of deep-sourced groundwater. In both cases, carbon sequestration by global groundwater circulation actively works to terminate surface habitability, rather than being a passive marker of warm Earthlike conditions. *Curiosity* will soon be in a position to discriminate between these two hypotheses. Our work links Mars sulfate cation composition, carbon isotopes, and climate change.

## 1. Introduction.

Liquid water flowed over the surface of Mars billions of years ago, and aqueous minerals also formed kilometers below the surface (e.g., McLennan & Grotzinger 2008, Ehlmann et al. 2011). However, the extent of hydrologic coupling between the surface/near-surface and deep subsurface on Early Mars is unknown. In one view, continuous permafrost isolated the deep hydrosphere from the surface, with only local and transient exceptions (e.g., Fastook & Head 2015, Fairén 2010, Schwenzer et al. 2012). In another hypothesis, surface and deep-subsurface waters repeatedly swapped places as part of a prolonged global groundwater cycle – vertical integration enabled by >10$^{7-8}$ yr of annual-averaged surface temperatures above the freezing point (Andrews-Hanna et al. 2010). The existence and extent of global deep-groundwater cycling are key unknowns for Early Mars climate and global hydrology (Wordsworth 2016), water loss (Usui et al. 2015), and habitability (Onstott et al. 2019). Moreover, global deep-groundwater flow could piston atmospheric $CO_2$ into the deep subsurface and fix it as deep carbonates. Uncertainty in the size of the deep carbonate reservoir is a major uncertainty in Mars' $CO_2$ evolution (e.g. Jakosky & Edwards 2018, Jakosky 2019).

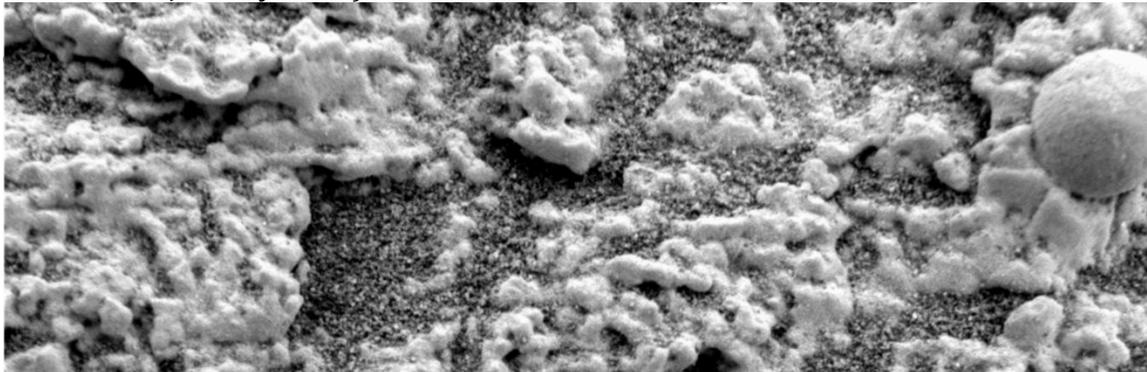

**Fig. 1.** Evidence for shallow-diagenetic alteration of the Burns Formation, Meridiani, Mars. Image is ~3 cm across. Laminae separated by <1 mm have different degrees of cementation. Halo girdles $Fe_2O_3$ concretion at right. Subframe of 1M130671710EFF0454P2953M2M1, Opportunity rover Microscopic Imager, sol 28. (After Grotzinger et al. 2005).

Central to the question of vertical isolation versus vertical integration of Early Mars' hydrosphere is the archive of ancient hydrology contained within mostly Hesperian-aged (3.6-3.2 Ga, based on crater chronology) sulfate-bearing sedimentary rocks (Malin & Edgett 2000, Bibring et al. 2007). This rock type was ground-truthed by the Mars Exploration Rover *Opportunity* investigation of the Burns Formation at Meridiani (e.g. Squyres et al. 2006). At Meridiani, texture and mineralogy record multiple stages of diagenesis – involving acid and oxidizing near-surface groundwater (Fig. 1) (McLennan & Grotzinger 2008). Burns Formation sandstones are ~40 wt% {Mg,Fe,Ca}$SO_4$. The cations for the sulfates were initially interpreted to be derived from slow wicking-to-the-surface and evaporation of deep-sourced saline groundwater, as part of a global groundwater circulation that could also explain the low elevations of the sulfate-bearing rocks (Andrews-Hanna et al. 2010). The rocks must have been altered in pH = 2-4 waters in order to explain the detection of jarosite (Tosca et al. 2008). The jarosite presumably formed near the surface. There, $Fe^{2+} \rightarrow Fe^{3+}$ oxidation (by UV photons, or by atmospheric $O_2$), or volcanic $H_2SO_4$, could drive acidity (Tosca et al. 2008, Baldridge et al. 2009, Hurowitz et al. 2010, Xie et al. 2017). *Opportunity*'s mapping of sulfate-bearing sedimentary rocks that experienced acid and oxidizing alteration (driving the formation of jarosite and hematite) has been extended by orbital surveys. Sulfate-bearing sediments (>$10^6$ km$^3$ in total) extend across much of Meridiani (5°S-10°N); Valles Marineris and nearby chaos (0-15°S); and Gale crater (5°S), among other sites (Bibring et al. 2007, Hynek & Phillips 2008, Grotzinger & Milliken 2012). Stratigraphic thicknesses of >2 km suggest millions of years of deposition. Thus, if groundwater circulation was the engine of sulfate formation, then the sulfate-bearing sedimentary rocks are evidence for millions of years of warm Earthlike conditions on Early Mars.

However, the Hesperian global deep-groundwater circulation hypothesis is disputed. For example, the sulfates themselves have been proposed to form at 220-270K by acid-rock reactions (Niles & Michalski 2009, Niles et al. 2017). The textual evidence for shallow flow of groundwater (Fig. 1) might be explained by a shallow, local source of water, such as seasonal meltwater (Kite et al. 2013a). The survival (within sedimentary rocks) of fragile minerals that would have been dissolved by high water/rock ratio alteration at depth argues against persistent circulation (e.g. Dehouck et al. 2017, Phillips-Lander et al. 2019). The low variation of K/Th (5300±220) on Mars' surface at 300km scales is an independent argument against global groundwater circulation (Taylor et al. 2006). Potassium is much more mobile than thorium, and so extensive aqueous alteration of the crust would be expected to enhance K at evaporation zones. But this is not observed (Taylor et al. 2006). Therefore, the existence or otherwise of a global and deep groundwater circuit on Early Mars remains an open question. Nevertheless, there is abundant evidence for groundwater movement on Early Mars: mineralized fracture-fills including Ca-sulfate veins are widespread (Okubo & McEwen 2007, Yen et al. 2017). Moreover, de-watering drove sediment deformation during early diagenesis at some sites (Rubin et al. 2017), and water was released from the subsurface to form some chaos terrains and associated outflow channels (Carr 2006). However, the duration and cause of all these flows remains unclear. In summary, the sulfate-bearing sedimentary rocks were altered by shallow groundwater (McLennan et al. 2005), but there is no consensus as to whether or not Hesperian Mars had a global hydrologic cycle including deep groundwater.

## 2. Global groundwater circulation implies carbonate sequestration if recharge waters equilibrated with the atmosphere.

Here, we test the global deep-groundwater hypothesis for formation of the Hesperian sulfates. We do this by tracing the hypothesis' implications for carbon sequestration and thus global atmospheric and climate evolution (Fig. 2). Our starting point is that $CO_2$-charged water is out-of-equilibrium with the basaltic crust of Mars. As a result, initially atmospherically-equilibrated water percolating along the long flow paths entailed by the global deep-groundwater hypothesis (>$10^3$ km; Andrews-Hanna et al. 2010) should deposit C as carbonate (e.g. Griffith & Shock 1995, Niles et al. 2013, Tomkinson et al. 2013, Melwani Daswani et al. 2016). Carbonate formation is inevitable for very long and deep flow paths through basalt, and mechanisms that could inhibit carbonate formation in the surface/near-surface (e.g. Bullock & Moore 2007) do not apply here. (Even for relatively short and shallow groundwater flow-paths, carbonate precipitation from groundwater has occurred on Mars: e.g. van Berk et al 2012, Ruff et al. 2014). Once emplaced deep within basalt, Hesperian carbonates should persist. That is because (unlike Earth) post-3.6 Ga Mars lacked a mechanism to heat carbonates to drive off $CO_2$, such as plate tectonics or global volcanism (Ogawa & Manga 2007), except locally (Glotch & Rogers 2013). Therefore, deep groundwater circulation between the atmosphere/surface and the basaltic crust of Mars implies one-way geologic sequestration of $CO_2$ (Niles et al. 2013) (Fig. 2).

By estimating the magnitude ($C_{seq}$) of the implied Hesperian $CO_2$ sequestration, we can test the global groundwater hypothesis. Possible outcomes of our test are:

- $C_{seq}$ > 10 bar: If $C_{seq}$ exceeds Mars' total estimated outgassed $CO_2$ inventory (1-10 bar; Stanley et al. 2011, Lammer et al. 2013), then the deep-global groundwater hypothesis and our understanding of Mars' composition are not consistent with one another.
- $C_{seq}$ > 1 bar: If $C_{seq}$ exceeds the $CO_2$ in the atmosphere at 3.6 Ga plus the amount of $CO_2$ outgassed 3.6-3.2 Ga (a subset of Mars' total estimated $CO_2$ inventory, because of pre-3.6 Ga $CO_2$ loss; Kite et al. 2014), then the deep-global groundwater hypothesis predicts a very thin atmosphere. In that case, the deep-global groundwater hypothesis is not consistent with the widely-held belief (Haberle et al. 2017) that an atmospheric pressure >0.1 bar is needed for extensive liquid water at the surface of Early Mars.
- $C_{seq}$ < 1 bar: If $C_{seq}$ is <1 bar, then there is no tension between the deep-global groundwater hypothesis and our understanding of Mars history. However, $C_{seq}$ > 0.1 bar would imply atmospheric drawdown and thus climate change during the time of deposition of the sulfate-rich sedimentary rocks (if the deep-global groundwater hypothesis is correct).

## 3. Method.

Unfortunately, $C_{seq}$ is poorly constrained by existing observations. Although crustal carbonates are observed in Mars meteorites (Bridges et al. 2001), and from Mars orbit (Wray et al. 2016), the quantity of crustal C on Mars is very uncertain. (The source of C for the crustal carbonates is often also uncertain - C degassed from intrusions, as well as C leached from crystalline rocks, are alternatives to drawdown of atmospheric C). Carbonate is not detected at most locations on Mars. If (on this basis) we were to assign an upper limit of 1 wt% carbonate to Mars' topmost 10 km of crust, then the corresponding upper limit on sequestration would be ~4 bars. 4 bars exceeds Mars' total estimated outgassed $CO_2$ inventory (Lammer et al. 2013). Therefore, currently published data provide little guidance on the partitioning and fate of Mars C.

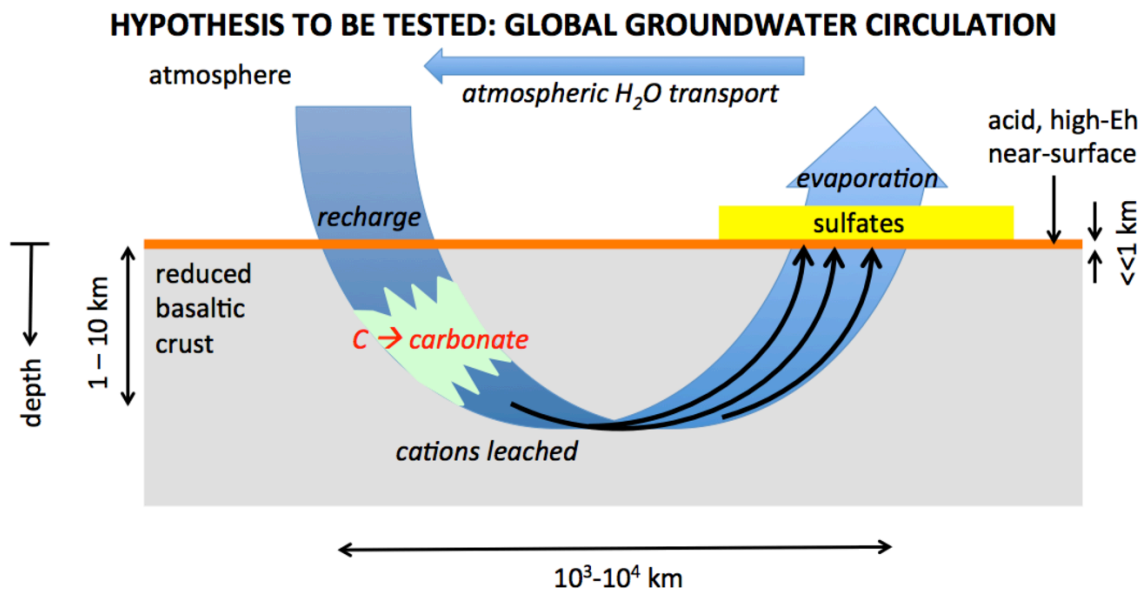

**Fig. 2.** The prevailing view of sulfate origin, to be tested here. Formation of sulfates from deep-sourced cations implies C sequestration by carbonate formation.

Therefore, we use a fluid-centered flow-through geochemical model (CHIM-XPT; Reed et al. 1998), combined with global mass balance, to test the deep-groundwater hypothesis by finding $C_{seq}$. This test requires answers to two sub-questions:

(1) How much water is needed to make the Hesperian sulfate-bearing rocks (§4.1)?

(2) How much C is sequestered per unit water? (How much of the initial C dissolved into water at the recharge zone goes into carbonate, and how much survives to reach the upwelling zone)? (§4.2)

For our CHIM-XPT calculations, our basalt (aquifer-host) rock composition is based on Mars rover measurements of fresh Mars basalt (McSween et al. 2006) and is as follows: 46.22 wt% $SiO_2$; 10.88 wt% $Al_2O_3$; 2.14 wt% $Fe_2O_3$; 17.08 wt% FeO; 0.44 wt% MnO; 10.49 wt% MgO; 8.35 wt% CaO; 2.66 wt% $Na_2O$; 0.11 wt% $K_2O$; 0.64 wt% $P_2O_5$; 0.84 wt% FeS; and 0.15 wt% Cl. To take account of the possibility of incongruent basalt dissolution (Milliken et al. 2009), we also considered a pure-olivine composition. Our olivine composition is also based on Mars-rover measurements for little-altered basalt (McSween et al. 2006) and is as follows: 34.75 wt% $SiO_2$; 42.39 wt% FeO; 22.86 wt% MgO (zero CaO)[1]. Precipitation of many minerals is suppressed for kinetic reasons. We do not explicitly consider Mars upper crust permeability in our model. The mean value of permeability is currently unknown, and estimated to lie in the range ($10^{-11}$ - $10^{-15}$) m² (by Hanna & Phillips 2005) for Mars' upper crust. For Mars' upper crust, permeability likely varies by orders of magnitude regionally (Harrison & Grimm 2009), and likely also varied with time on Hesperian Mars due to the competition between fracture-sealing processes (such as carbonate mineral precipitation within fractures), and fracture creation (by tectonics, impacts, and cracking to accommodate magmatic intrusions) (Sleep & Zoback 2007). In effect we assume permeability is high enough that permeability is not limiting for deep, global groundwater circulation. Thus, our upper bounds on the vigor of deep global groundwater circulation only become stronger if we are wrong about permeability. Our runs assume an atmosphere with minor $O_2$ (as for present Mars), but running with zero $O_2$ would not affect the conclusions. That is because any dissolved oxygen is quickly consumed at W/R < $10^4$ by early FeO(OH) precipitation, and the fluid subsequently stays fairly reduced along the flow path. Runs are carried out at 0.01°C, which is thought to be appropriate for the near-surface of Early Mars. Details on our CHIM-XPT runs are contained in the Supplementary Information.

## 4. Results.

### 4.1. How much water is needed?
We use two methods to estimate the water demand if global-groundwater circulation was the engine of Hesperian sulfate formation. One method is based on jarosite stability (Hurowitz et al. 2010), and the other method is based on cation supply. The larger of the two is the relevant constraint for the hypothesis that deep-sourced groundwater supplied the cations for the sulfates at Meridiani. This turns out to be the cation supply constraint.

---

[1] In-situ X-ray diffraction data for residual Ol in Gale sediments is more Mg-rich (Morrison et al. 2018), as expected for water-altered sediments (Stopar et al. 2006).

Both methods require rock volume as input. The volume of sulfate-bearing sedimentary rock on Mars is >$3.4 \times 10^5$ km$^3$ today (Niles & Michalski 2012, Hynek & Phillips 2008). The outcrops have eroded surfaces, so were once more voluminous. Moreover, the layers in the sulfates were originally very close to flat in the deep-groundwater hypothesis, in contrast to the ~10° topographic slopes and ~5° layer dips measured for outcrops today (Kite et al. 2016). Thus, in the deep-groundwater hypothesis, erosion has removed (87-92)% of the original volume of sulfate-bearing sedimentary rocks, which is implied to be 2.9-4.0 × 10$^6$ km$^3$ (Andrews-Hanna et al. 2010, Andrews-Hanna 2012, Michalski & Niles 2012, Zabrusky et al. 2012). (This entails a very large pre-erosion S content. It is unclear where this S could have been sourced from, and where it could be hidden today; Michalski & Niles 2012).

<u>Jarosite-stability method.</u> The water/rock ratio (kg/kg) for the Burns Formation was $10^2$-$10^4$, based on the observation of jarosite (Tosca et al. 2008, Tosca & McLennan 2009, Hurowitz et al. 2010) [2]. In this model, pH is lowered by acid produced by dissolving rock, and then oxidizing rock-sourced $Fe^{2+}$ (Hurowitz et al. 2010). Jarosite will not form, and $Fe^{3+}$-copiapite or rhomboclase will form instead, if the water/rock ratio is <$10^2$ (Tosca & McLennan 2009) – but jarosite is in fact observed. The implied Burns Formation water/rock ratio of $10^2$-$10^4$ is (3-300)× that of Andrews-Hanna et al. (2010), who assumed a salinity of 80% that of seawater (i.e., W/R ~ 40). For rock density 2500-2800 kg/m$^3$ we get a total water demand of $5 \times 10^{20}$ – $5 \times 10^{22}$ kg for the sulfate rocks observed today, i.e. 3-300 km Global Equivalent Layer of water. (This assumes that the bulk of the Burns Formation materials interacted with water, consistent with the paucity of olivine in the Burns Formation). This increases to 30-3000 km taking account of the now-eroded sulfates entailed by the groundwater hypothesis. Even the lower bound is probably more than can be stored in surface-exchangeable reservoirs in the crust at any one time (Clifford & Parker 2001). Therefore, this quantity of water strongly suggests a hydrologic cycle, but does not constrain whether or not this cycle included the deep subsurface.

<u>Cation-supply method.</u> A second, independent constraint on the water demand comes from basalt-water equilibration as calculated using CHIM-XPT simulations (Fig. 3). The CHIM-XPT simulations output dissolved-cation content, dissolved-C content, and pH, all as a function of water/rock ratio. To select Mars-relevant output, we use terrestrial basaltic aquifers as a guide to pH, and then interpolate in the CHIM-XPT output as a function of pH to obtain the corresponding dissolved-cation concentration, as well as what proportion of C has been precipitated from fluids into carbonate minerals.[3]

---

[2] Throughout this paper, we distinguish between the water:rock ratio computed by dividing the time-integrated water flux by the mass of sulfate-rich rock, and the water:rock ratio of the outlet fluid (which is a measure of the extent of equilibration between groundwater and basalt for each water parcel that upwells to the surface). For example, a water:rock ratio of 1 in CHIM-XPT has total dissolved solids typically ~0.1 mol/liter, so to build up a sulfate deposit containing 10 wt% deep-groundwater-supplied-cations would require a time-integrated water/rock ratio $\gg 1$.

In order to relate groundwater cation concentrations (mol/liter) output from CHIM-XPT simulations to the water demand (liter) for the sulfate-rich sedimentary rocks, we need to know the cation content (moles) of the sulfate-rich sedimentary rocks. We estimate this in two ways.

a) *Ground-truth approach:* We use Burns formation measurements as a proxy for sulfate composition (see §5.1 for discussion). According to Squyres et al. (2006), the Burns formation is a mixture of (i) a siliciclastic component that was leached of 55% (by moles) of its divalent cations and (ii) a subsequently-added evaporitic sulfate component; this mixture was subsequently diagenetically modified. This implies that >3.4 wt% of the Burns formation (the Fe in jarosite according to McLennan & Grotzinger 2008) consists of groundwater-transported Fe, rising to 6.5 wt% for a more involved calculation[4]. However, Fe is very insoluble in reducing, circumneutral-to-alkaline waters, such as basalt-equilibrated aqueous fluids. The equilibrium concentration of Fe in the fluid is <0.1 mmol/liter in CHIM-XPT for pH > 7 and $pCO_2$ in the range 0.006 bar to 6 bar, i.e. ≲6 ppmw (Fig. 3). $pCO_2$ ≥0.2 bar is favored for Hesperian Mars, according to climate models (Haberle et al. 2017). This gives a water demand of $\sim 10^{23}$ kg, i.e. >500 km GEL of water, for the >4 × $10^5$ km$^3$ of sulfate-bearing sedimentary rock seen today. (A similar argument applies to $Ca^{2+}$, which is buffered to <0.04 mmol/liter in our 0.2 bar basalt calculation, and <0.003 mol/liter in our 2-bar calculation.)

b) *Orbital-spectroscopy approach:* Orbital near-infrared spectroscopic data for sulfate-rich rocks outside Meridiani show only occasional evidence for Fe-sulfates, with Mg-sulfates being much more important (e.g. Gendrin et al. 2005, Murchie et al. 2009, Wang et al. 2016). Therefore, we consider the $MgSO_4 \bullet nH_2O$ cation endmember, with *n* = 2 to represent a mix of starkeyite and singly-hydrated Mg-sulfate (Wang et al. 2016). We assume Mg-sulfates make up (40±20)% of the mass of the sulfate-rich rocks. The corresponding water demand constraint is much looser for Mg-sulfates (i.e., orbital spectroscopy approach) than for Burns-formation composition (i.e., ground truth approach). This is because [Mg] can reach 0.03 mol/liter (i.e. 0.7 g/kg) for the highest $pCO_2$ levels we investigate (Fig. 3).

---

[3] We do not use the terrestrial basalt aquifer cation and C data directly, in part because Early Mars is thought to have had atmospheric $pCO_2$ much greater than that of modern Earth.

[4] According to McLennan (2012), Burns formation rock has mean [$SiO_2$] of 37.0 wt%. We look up 37.0 wt% $SiO_2$ in the Sulfur-Plus-Cations (S Addition) worksheet of Data Set S2 (jgre21007-sup-0003-2018JE005718-ds02.xlsx) of McCollom (2018). This gives that 53% of Burns formation Fe consists of groundwater-transported Fe, i.e. 6.5 wt% of the rock. The tab in McCollom (2018) is a quantification of the argument in Squyres et al. (2006). However, Hurowitz & Fischer (2014) state that a lower degree of alteration is consistent with the basic hypothesis of Squyres et al. (2006). We use 3.4 wt% (Fe in jarosite) as a lower bound, and 6.5 wt% as an upper bound.

## 4.2. How much C is drawn down per unit water?

The C that is drawn down is the C dissolved in the water at the recharge zone, minus C that reaches the outlet. The difference is due to carbonate formation.

*C dissolved in the water at the recharge zone.* For water-$CO_2$ equilibration at 0.01°C, CHIM-XPT outputs 0.39 mol/liter $CO_{2(aq)}$ at $pCO_2$ = 6 bar, decreasing near-linearly with $pCO_2$. Thus at 0.6 bar partial pressure, 1.7 $CO_2$-equivalent-g/l goes into the water at the recharge zone. This corresponds to 3-5 bars of $CO_2$ for the 300-to-500-km-thick Global Equivalent Layer of evaporated water that was calculated in §4.1.

*C sequestered along the flow-path.* What fraction of the initial C in the water in the recharge zone goes into carbonate? CHIM-XPT output shows that C decreases with increasing distance along the flow-path, and the amount of sequestration depends on the pH of the water (Fig. 3). Observations (fieldwork and experiment) show pH > 7 for basalt-buffered waters. pH is 9-10 for groundwaters in basalt-buffered aquifers in Iceland, which is consistent with the pH of basalt-buffered fluids in the laboratory when isolated from the atmosphere (Arnórsson et al. 2002). pH for thermal springs in the basaltic Deccan Traps is 8.2±0.5 (Minissale et al. 2000). If the aquifer rock dissolves non-congruently, with preferential dissolution of olivine, then pH = 8-11 is expected (Kelemen et al. 2011). Drawdown of $CO_2$ as carbonates in basaltic aquifers has been documented in Iceland (Flaathen et al. 2009), among other locations. In seafloor hydrothermal systems on the flanks of mid-ocean ridges, 80-90% of recharging seawater C can be sequestered as carbonate (Walker et al. 2008).

These results (and their relevance to Mars) can be understood in terms of a ratio of timescales. The flow-through time for a large basaltic aquifer is compared to the dissolution time for rock adjacent to the flow. If the flow time is much more than the dissolution time (i.e. Damköhler number $Da \gg 1$), then the water will be equilibrated with the rock when it reaches the outlet. Darcy's law gives $u = k/\mu \times \Delta p/\Delta x$, where k is the permeability, and $\mu = 10^{-3}$ Pa s is dynamic viscosity of water. Pressure head $dp = \rho \times g \times \Delta z = 10^3$ kg/m$^3$ × 3.7 m/s$^2$ × $10^4$ m = 4 × $10^7$ Pa (here we have conservatively chosen a large $\Delta z$), and flow path length $\Delta x = 10^7$ m. If we let k = $10^{-12}$ m$^2$, then we obtain $u$ = 0.01 m/yr. The cross-sectional area of fractures through which the fluid is moving is maybe 1% of the rock volume ($\ll$ the porosity), so the speed of the water is ~100× faster. The flow-through time is $\Delta x/100u$ = $10^7$ yr. $10^7$ yr is long enough to dissolve mineral grains (Milliken et al. 2009), and thus to allow minerals and water to reach equilibrium. Thus, as stated in Andrews-Hanna et al. (2010), for "flow timescales on the order of millions of years […] there is ample time for these fluids to equilibrate with the aquifer matrix." Thus, deep groundwater should have pH 8.2±0.5 (or even larger) and ≳ ½ of input C will be sequestered as carbonate (Fig. 3) – i.e., many bars.

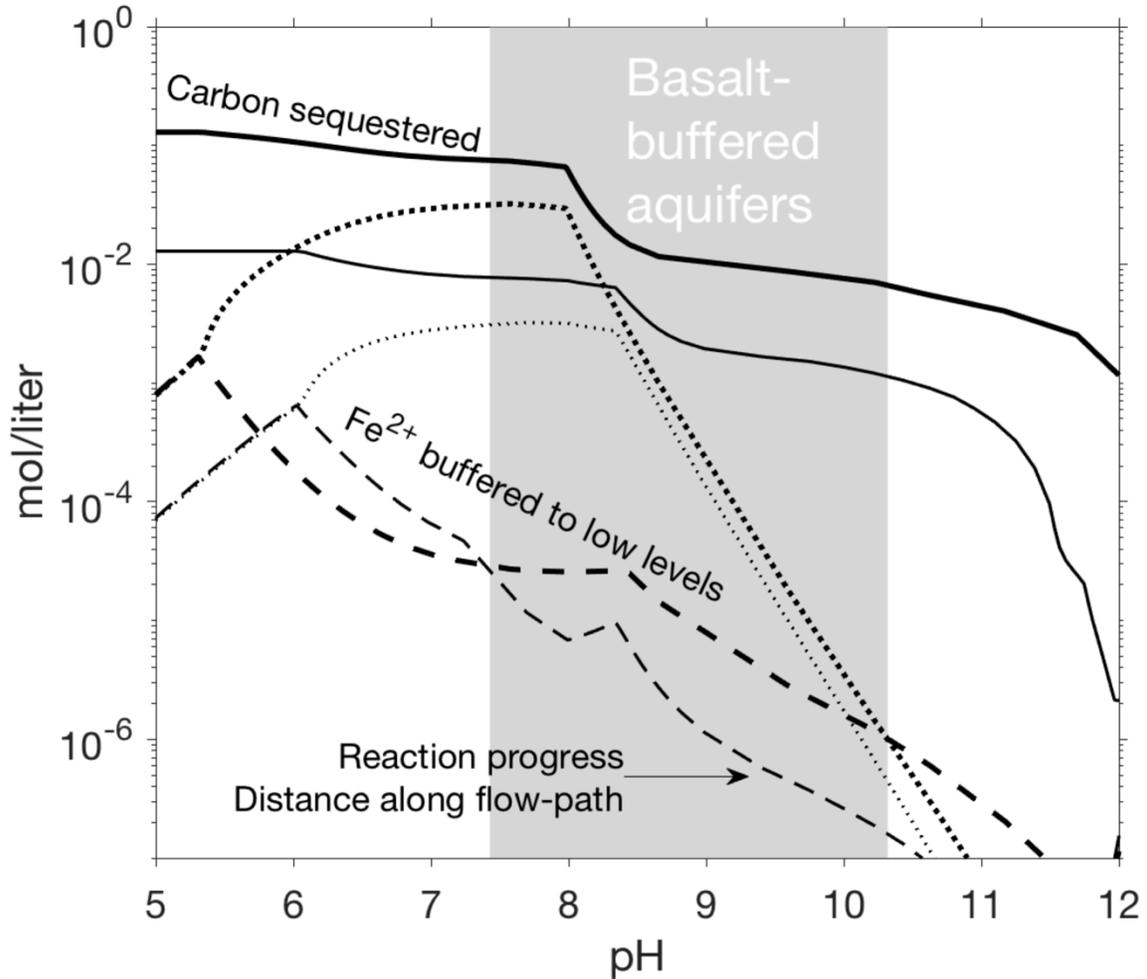

**Fig. 3.** CHIM-XPT output for water-basalt interaction for 0.2 bars (thin lines) and 2 bars pCO$_2$ (thick lines). Solid lines show total C in aqueous fluid; dashed lines show Fe in fluid; and dotted lines show Mg in fluid.

### 4.3. Monte Carlo procedure.

The estimates in §4.1 and §4.2 are only rough estimates because many parameters are uncertain. To take account of this uncertainty, we used a Monte Carlo approach. First, to take account of the possibility of incongruent basalt dissolution, we considered dissolution of Mars basalt, and also olivine-only dissolution, as endmembers (Milliken et al. 2009). $C_{seq}$ for these two endmembers is plotted separately in Fig. 4. For each of these two cases, we varied other relevant parameters, as follows. (1) Uncertainty in the original volume of sulfate-bearing sedimentary rocks $V_{sed}$ entailed by the global-groundwater hypothesis. We adopt a uniform prior in the range (2.9-4.0) × 10$^6$ km$^3$. This sums the present-day outcrop volume of >3.4 × 10$^5$ km$^3$ (Niles & Michalski 2012, Hynek & Phillips 2008); the eroded volume of 0.9-1.7 × 10$^6$ km$^3$ entailed by the global groundwater hypothesis for Meridiani (Zabrusky et al. 2012); and the eroded volume of (>1.64)-2 × 10$^6$ km$^3$ entailed by the global groundwater hypothesis for Valles Marineris (Andrews-Hanna 2012, Niles & Michalski 2012). (2) Uncertainty in the sedimentary-rock density $\rho_{sed}$ (uniform prior in the range 2500-2800 kg/m$^3$). (3) Uncertainty in the percentage of rock mass corresponding to added

Fe ($Fe_{added}$), ranging from 3.4-6.5 wt% (log-uniform prior, i.e. conservatively favoring smaller values). (4) pH was randomly chosen from the pH measured from 25 thermal springs in the Deccan Traps (Minissale et al. 2000) and 80 low-ground springs in Iceland (Arnórsson et al. 2002). We weighted the random sampling so that 50% of trials used a Deccan-basaltic-aquifer pH, and 50% of trials used an Iceland-basaltic-aquifer pH. (5) $pCO_2$ was varied randomly between values of 0.2, 0.6, and 2 bars. We use 0.2 bar as our lower limit; $pCO_2$ < 0.6 bar will lead to a frozen Mars surface according to existing models (e.g. Haberle et al. 2017). For the jarosite method, we also considered (6) uncertainty in the initial W/R, with a log-flat prior between limits of $10^2$ and $10^4$. (Our calculations do not take into account the possibility of surface temperatures ~20°C on Early Mars, which would reduce both dissolved $CO_2$ concentrations, and carbonate-mineral solubility). With these assumptions,

$$C_{seq} \times 10^5 = (\,[C]^{inlet}_{(pH,\,pCO2)} - [C]^{outlet}_{(pH,\,pCO2)}\,) \times W_i \times 0.044 \text{ kg/mol} \times 3.7 \text{ g}$$

where $C_{seq}$ is in bars, and water demand $W$ can correspond to any of our three methods for estimating water demand (§4.1),

$$W_{jarosite} = (W/R)\,\rho_{sed}\,V_{sed}$$

$$W_{Fe} = Fe_{added}\,\rho_{sed}\,V_{sed}\,/\,[Fe]_{pH,\,pCO2}$$

$$W_{Mg} = Mg_{added}\,\rho_{sed}\,V_{sed}\,/\,[Mg]_{pH,\,pCO2}$$

where $[Fe]_{pH,\,pCO2}$ and $[Mg]_{pH,\,pCO2}$ are from calculations like those shown in Fig. 3.

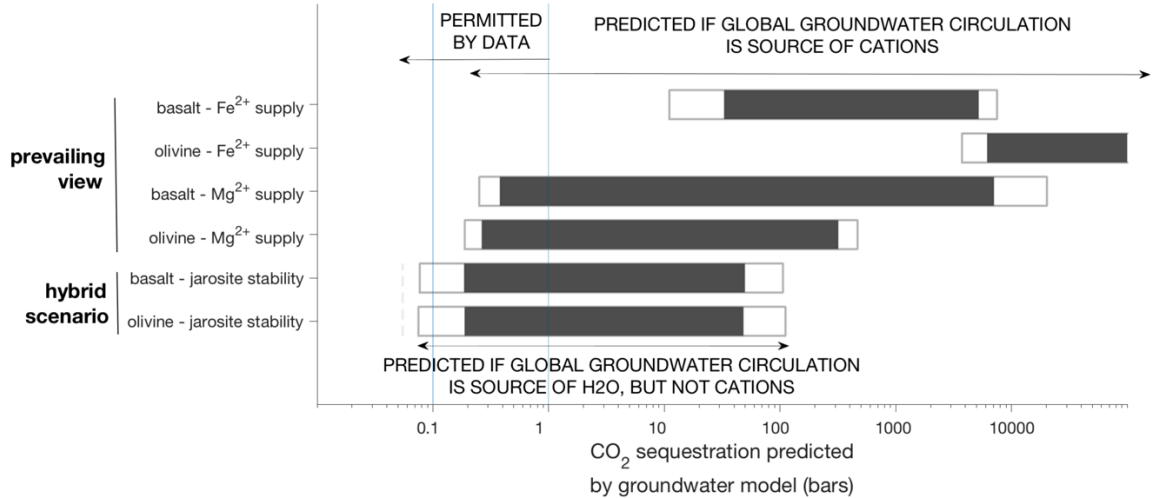

**Fig. 4.** $CO_2$ sequestration. The dark gray bars show the 90% range of uncertainty and the light gray boxes show the 99% range of uncertainty, both as output from the Monte Carlo procedure described in the text. The blue line shows the upper limit of outgassed $CO_2$. The thin dashed gray lines show the lowest values permitted by the jarosite stability calculation.

We also plot the "minimum carbonate drawdown" (gray dashed line in Fig 4). For this we assumed a pH equal to the minimum measured for thermal springs from Deccan by Minissale et al. (2010), i.e. 7.43, and a $pCO_2$ at the bottom of our range (0.2 bar).

Using the ground-truth approach (i.e. treating *Opportunity* data as representative of sulfate-rich Hesperian sediments), the results show enormous $CO_2$ drawdown in order to match $Fe^{2+}$-supply constraints (Fig. 4). Indeed, $CO_2$ drawdown is much greater than plausible Hesperian $CO_2$ sources, beginning-of-Hesperian $pCO_2$, or the sum of the two (Stanley et al. 2011, Kite et al. 2014). The implication is that Mars' atmosphere would have been driven underground. Since an atmosphere is required in order to warm the climate enough to prevent formation of a global cryosphere and thus allow global groundwater circulation, this hypothesis under-mines itself. However, this result does not by itself disprove the hypothesis of global groundwater circulation. For example, using the orbital spectroscopy approach (i.e. approximating the sulfate component of the sulfate-rich rocks as being exclusively Mg-sulfates), the $CO_2$ drawdown is much less (Fig. 4). The 90% range of uncertainty for $CO_2$ drawdown includes some that are small enough to be consistent with existing data. There are other workarounds and alternatives, as we now discuss.

## 5. Discussion

### 5.1. Rescue for the global-groundwater hypothesis?

There are several work-arounds for the global groundwater circulation hypothesis that can improve the agreement between $pCO_2$ constraints and $CO_2$ drawdown. We list these below, from the least likely to (in our judgement) the most likely:

*(a) Overestimated $V_{sed}$?* This work-around posits that the volume of sulfate-bearing outcrop seen today is not much less than the pre-erosion volume. This may be true (e.g. Niles & Michalski 2012, Kite et al. 2013b), but is hard to square with the hypothesis that the engine of formation for sulfates was a global groundwater circulation. Suppose that the present-day mounds formed via upwelling of deep-sourced groundwater and had a maximum volume not much more than their present-day volume. Then, given the topographic isolation of the present-day mounds, the corollary is that the mounds are spring mounds. This is unlikely based on structural geology analysis (e.g. Kite et al. 2016).

*(b) The atmosphere at the start of the Hesperian was very thick, and/or volcanoes outgassed >10 bars of $CO_2$ during the Hesperian.* This is unlikely. For example, lab experiments, Mars-meteorite-based redox estimates, and geologic constraints on the volume of post-3.6 Ga magmatism, suggest only <0.1 bar was released into the atmosphere during the Hesperian (Stanley et al 2011, Grott et al. 2011).

*(c) Waters did not equilibrate with the Hesperian atmosphere.* In this picture, basal melting of glaciers recharges the aquifers. Because ice traps <100 ppmw air, the basal ice-melt water holds little C. This hypothesis is in tension with models of ice flow on Early Mars, which predict little or no basal melt (Kite & Hindmarsh 2007, Fastook & Head 2015). Another way to avoid equilibration between groundwater and the Hesperian atmosphere is "one-shot" upwelling-from-depth of very saline groundwater. Highly saline waters could be sourced by dissolution of buried Noachian sulfates (Zolotov & Mironenko 2016), or infiltration of water from a primordial ocean.

*(d) Later waters flowed through fractures coated with carbonate precipitated from earlier waters, so that later waters arrived at the evaporation zone without having equilibrated with basalt.* This work-around predicts that waters upwelling at Meridiani would have been equilibrated with carbonate. If so, carbonate precipitation would have occurred at Meridiani. However, carbonate is not observed in the Burns formation, so this workaround is unlikely.

*(e) Carbonate recycling by acid weathering, or thermal breakdown of carbonates by heat from lava.* In this work-around, carbonates do form, but at depths shallow enough for subsequent dissolution by volcanogenic $H_2SO_4$ or (in the case of Tharsis recharge) by heat from overlying lava (Glotch & Rogers 2013). This recycles $CO_2$ back into the atmosphere.

*(f) Rocks seen by Opportunity have elemental compositions that are not representative of Hesperian sulfates on Mars.* (This is essentially the "Orbital spectroscopy" approach, corresponding to the $Mg^{2+}$ supply constraint in Fig. 4). Orbiter infrared spectroscopy shows association of Fe-oxides and sulfates (e.g. Bibring et al. 2007) – the "Laterally Continuous Sulfate" facies of Grotzinger & Milliken (2012) which includes much of Meridiani, Valles Marineris and nearby chaos, and Gale crater, among other sites. However, Mars orbiter visible-and-near-infrared spectroscopy cannot precisely constrain bulk rock Fe content, and Mars orbiter gamma ray spectroscopy does not resolve Hesperian sulfate-rich bedrock outcrops. Fortunately, *Curiosity* will soon arrive at sulfate-rich rocks in Gale, 8000 km from Meridiani, and test the hypothesis that sulfate-rich Hesperian rocks are similar around the planet. If the Burns formation is unusually rich in Fe-sulfates and secondary Fe-oxides, then the global-groundwater circulation hypothesis predicts that the dominant cation transported is Mg-sulfate, with 100× smaller concentrations of Fe and Ca (Fig. 3; Wang et al. 2016).

*(g) Waters from the deep subsurface were not a major source of cations for the sulfates; instead, cations were leached from wind-blown siliciclastic material. Thus, cation vertical transport distance in aqueous solution (including late-stage remobilization, e.g. of $MgSO_4$) was $\ll 1$ km. Deep-sourced groundwater is not the source of the extra cations in the "added sulfates" at Meridiani or elsewhere.*
We refer to this work-around as the "hybrid scenario". It corresponds to the lowest two rows in Fig. 4, and the central panel in Fig. 5. In this work-around, almost none of the cations for the sulfates are derived from deep-sourced groundwater. Instead, deep-sourced groundwater provides only a small fraction of the $Fe^{2+}$ (or $Ca^{2+}$), but almost all of the $H_2O$. This allows the water:rock ratio to be as low as compatible with jarosite stability; i.e., 100:1 (Hurowitz et al. 2010). In the hybrid scenario, Fe oxidation creates acidity that may have allowed modest additional leaching of rocks immediately beneath the Burns formation (e.g. the Shoemaker formation). In the hybrid scenario, the build-up of the evaporitic sandstone is rate-limited by aeolian processes. Without wind-blown material, there are almost no cations and thus almost no sulfates. If this is true, then no "clean" sulfate evaporites will be found on Mars: in other words, Martian sulfates are necessarily dirty[5]. This can be tested by *Curiosity*'s imminent exploration of sulfate-rich rocks at Gale crater.

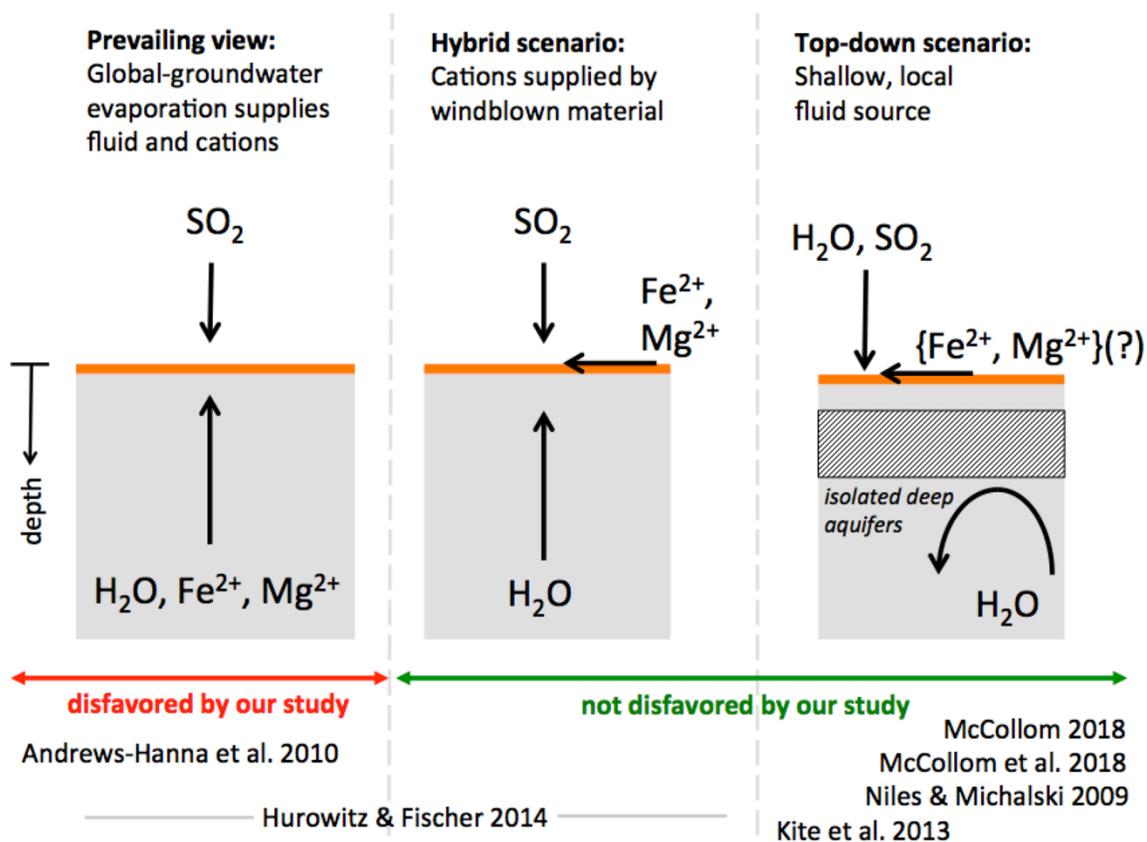

**Fig. 5.** Summaries of 'prevailing view,' 'hybrid scenario,' and 'top-down scenario' for formation of sulfates hosted in the sedimentary rocks of Hesperian Mars.

Hybrid scenario (g) permits the global groundwater circulation hypothesis to match data. However, if the deep-sourced groundwater is not the source of the salinity in the evaporites at Meridiani, then there is no longer a geochemical rationale for appealing to deep-sourced groundwater. Water on Mars does not have to come from below, as we now discuss.

**5.2. Alternatives to the global-groundwater hypothesis.**

Sulfates can form by reaction between sulfuric acid and olivine at ultracold temperatures (e.g. Niles & Michalski 2009, Niles et al. 2017). The possibility that this process was the engine of sulfate formation at Meridiani is not ruled out by bulk geochemical data. Indeed, it has been proposed that Burns formation bulk geochemistry can be explained by S/$SO_2$ addition only, with no need for extra cations (McCollom 2018, and references therein). Addition of S/SO$_2$ alone, without cation addition, accounts for the lack of evidence at Meridiani for Na, K, and Si mobility. This lack of mobility is inconsistent with the expectation that these elements should be mobilized along with Fe and Mg (McCollom 2018; and our CHIM-XPT output, see Supplementary Information). This S-added scenario is an alternative to the view of the Mars Exploration Rover team, which is that Burns formation bulk geochemistry is the result of addition of both S and cations (Squyres et al. 2006, Hurowitz & Fischer 2014, Cino et al. 2017). Indeed, the totality of the textural, compositional and stratigraphic data requires movement of groundwater at >1 m vertical and >100 m horizontal scale (Grotzinger et al. 2005, McLennan et al. 2005). This textural evidence is not incompatible with the idea that the sulfates themselves were formed by reaction between SO$_2$ and basalt: perhaps the bulk geochemistry is set by S/SO$_2$-addition, with later modification (hematite concretion formation, perhaps MgSO$_4$-mobilization) by groundwater. In future, instruments that tie geochemistry to texture at sub-mm scale will be useful to resolve the ambiguity of bulk geochemical analysis – e.g., the Planetary Instrument for X-ray Lithochemistry (PIXL) on Mars 2020 (Allwood et al. 2015). In the meantime, our calculations, by themselves, can be reconciled with both the salt-added and S-added views (Fig. 5). (In Fig. 5, only a single, atmospheric source for SO$_2$ is indicated, for simplicity. An atmosphere-derived contribution to Mars S is indicated by isotope data; Franz et al. 2017. FeS$_2$-derived S could also contribute to the sulfates; Dehouck et al. 2012.)

---

[5] With hindsight, this may explain the non-detection of a clean playa-lake deposit along *Opportunity*'s traverse.

Global groundwater circulation was originally proposed in part to resolve the recognized conundrum of sulfate build-up at Meridiani in the absence of a closed topographic basin. This conundrum is addressed in the global groundwater circulation model (Andrews-Hanna et al. 2010). Alternative solutions to the conundrum exist – one example is the "flypaper" model (Kite et al. 2013a). In the "flypaper" model, on $10^{8-9}$ yr timescales, windblown material is abundant, and migrates globally. Windblown sediments can undergo acid leaching and aqueous cementation, but only where surface water is available. Surface water is only available (according to the model of Kite et al. 2013a) in locations where snowpack can seasonally melt. Seasonal meltwater has an expected Gyr-integrated spatial distribution (for a Mars climate on the cusp of final dry-out) that is a good match for the observed spatial distribution of light-toned sedimentary rocks (Kite et al. 2013a). In this model, the spatial distribution of light-toned sedimentary rocks corresponds to zones of past seasonal snowmelt, and the snowmelt acted as "flypaper" that trapped windblown material.

### 5.3. Linking cation composition, carbon isotopes, and climate change.

It is interesting that our central estimate of atmospheric drawdown ($C_{seq}$) is so large, even for the hybrid scenario (Fig. 4). If global groundwater circulation did operate (Fig. 5), then it is likely that carbonate sequestration was the principal sink for CO$_2$ at times when global groundwater circulation was active during the Hesperian – faster than escape-to-space (Jakosky et al. 2018). This hypothesis might be tested at Mt. Sharp with *Curiosity* logging of $\delta^{13}$C. Measurements indicating a decrease in atmospheric $\delta^{13}$C during the interval of sulfate formation would be consistent with carbonate sequestration as the dominant sink for CO$_2$ during the Hesperian. This is because $^{13}$C is preferentially incorporated in the carbonate, and some of the lightened C survives to the zone of upwelling (Fig. 3).

Moreover, our analysis links *Curiosity* measurements of cation content at Mt. Sharp to the prevailing view that evaporation of water from global groundwater circulation provided both cations and water for the sulfates. *Curiosity* detection of Fe-sulfates or abundant Fe-oxides would further disfavor the prevailing view (Figs. 4-5), whereas *Curiosity* detection of Mg-sulfates with no secondary Fe-minerals could be consistent with the prevailing view (Fig. 4).

**6. Summary and conclusions.**

We quantify the "carbon tax" for a global hydrologic cycle including deep groundwater (recharge → deep aquifers → evaporation → recharge) on Hesperian Mars. We find that if deep-sourced groundwater is the source of the cations needed to explain the mineralogy of the sulfate-rich rocks, then the $CO_2$ sequestration is >0.3 bars (>30 bars if the Fe-contents measured by *Opportunity* ground-truth are representative of sulfate-rich Hesperian layered sediments). These are conservative estimates, because pH > 8 waters, expected for long flow paths, can transport even fewer cations per unit $CO_2$ sequestration. This increases the "carbon tax." Indeed, our central estimates of $CO_2$ sequestration (as carbonate) are all well in excess of available $CO_2$ (Fig. 4). This tension does not arise for alternatives to the global groundwater circulation hypothesis, such as a top-down water supply (e.g. Kite et al. 2013a). This tension also does not arise for a 'hybrid' scenario where groundwater provides dilute fluids, and cations are brought in by windblown material (Fig. 5).

All these $CO_2$-drawdown numbers are large enough to affect climate. Therefore, a global groundwater circulation that could extend Mars surface habitability on a drying planet would contribute to its own demise. The low end of this $CO_2$ sequestration range is not ruled out by the data, and predicts a Hesperian downward trend in δ$^{13}$C that is potentially testable by *Curiosity*. Our work does not disprove the hypothesis of global groundwater circulation on Hesperian Mars. Our results suggest potential problems with (and new tests for) the geochemical justification for the hypothesis of global groundwater circulation on Hesperian Mars.

**Acknowledgements:** Particular thanks to Joel Hurowitz for insightful comments on a draft, and to two anonymous reviewers for comments and corrections. We thank Bill McKinnon for editorial handling. We thank Paul Niles, Andy Heard, Mark Reed, Joe Michalski, Nick Tosca, and Reika Yokochi, for discussions, and Bruce Jakosky for sharing a useful preprint. This work was enabled and inspired by the magnificent Mars Exploration Rover *Opportunity* mission (2004-2018). Part of this research was supported by a NASA grant (NNX16AG55G) to the University of Chicago. M.M.D.'s portion of the work was done partly as a private venture and not in the author's capacity as an employee of the Jet Propulsion Laboratory, California Institute of Technology. All CHIM-XPT output files, and all scripts needed to reproduce the results presented here, can be obtained for unrestricted further use by contacting the lead author.

**References.**

Allwood, A., et al., "Texture-specific elemental analysis of rocks and soils with PIXL: The Planetary Instrument for X-ray Lithochemistry on Mars 2020," 2015 IEEE Aerospace Conference, Big Sky, MT, 2015, pp. 1-13.

Andrews-Hanna, J.C. et al., 2010. Early Mars hydrology: Meridiani playa deposits and the sedimentary record of Arabia Terra. J. Geophys. Res. 115, E06002.

Andrews-Hanna, J.C., 2012. The formation of Valles Marineris: 3. Trough formation through super-isostasy, stress, sedimentation, and subsidence. J. Geophys. Res. 117, E06002.

Arnórsson, Stefán; Gunnarsson, Ingvi; Stefánsson, Andri; Andrésdóttir, Audur; Sveinbjörnsdóttir, Árny E., 2002, Major element chemistry of surface- and ground waters in basaltic terrain, N-Iceland.: I. primary mineral saturation, Geochim. Cosmocim. Acta, 66, 4015-4046.

Baldridge, A. M.; Hook, S. J.; Crowley, J. K.; Marion, G. M.; Kargel, J. S.; Michalski, J. L.; Thomson, B. J.; de Souza Filho, C. R.; Bridges, N. T.; Brown, A. J., 2009, Contemporaneous deposition of phyllosilicates and sulfates: Using Australian acidic saline lake deposits to describe geochemical variability on Mars, Geophys. Res. Lett., 36, CiteID L19201.

Bibring, J.-P. et al., 2007. Coupled ferric oxides and sulfates on the martian surface. Science 317, 1206–1210.

Bridges, J. C.; Catling, D. C.; Saxton, J. M.; Swindle, T. D.; Lyon, I. C.; Grady, M. M., Alteration Assemblages in Martian Meteorites: Implications for Near-Surface Processes, Space Science Reviews, v. 96, 1/4, 365-392.

Bullock, M.A., Moore, J.M., 2007. Atmospheric conditions on early Mars and the missing layered carbonates. Geophys. Res. Lett. 34, 19201.

Carr, M.H., 2006, The Surface of Mars, Cambridge University Press.

Clifford, Stephen M.; Parker, Timothy J., 2001, The Evolution of the Martian Hydrosphere: Implications for the Fate of a Primordial Ocean and the Current State of the Northern Plains
Icarus, Volume 154, Issue 1, pp. 40-79.

Dehouck, E.; Chevrier, V.; Gaudin, A.; Mangold, N.; Mathé, P.-E.; Rochette, P., 2012, Evaluating the role of sulfide-weathering in the formation of sulfates or carbonates on Mars, Geochim. Cosmocim. Acta, 90, 47-63.

Dehouck, E.; McLennan, S. M.; Sklute, E. C.; Dyar, M. Darby, 2017, Stability and fate of ferrihydrite during episodes of water/rock interactions on early Mars: An experimental approach, J. Geophys. Res.: Planets, 122, 2, 358-382.

Ehlmann, Bethany L.; Mustard, John F.; Clark, Roger N.; Swayze, Gregg A.; Murchie, Scott L., 2011, Evidence for low-grade metamorphism, hydrothermal alteration, and diagenesis on Mars from phyllosilicate mineral assemblages, Clays and Clay Minerals, vol. 59, 4, 359-377

Fairén, Alberto G., 2010, A cold and wet Mars, Icarus, Volume 208, Issue 1, p. 165-175.

Fastook, James L.; Head, James W., 2015, Glaciation in the Late Noachian Icy Highlands: Ice accumulation, distribution, flow rates, basal melting, and top-down melting rates and patterns, Planetary and Space Science, 106, 82-98.

Flaathen, Therese K.; Gislason, Sigurður R.; Oelkers, Eric H.; Sveinbjörnsdóttir, Árný E., 2009, Chemical evolution of the Mt. Hekla, Iceland, groundwaters: A natural analogue for CO2 sequestration in basaltic rocks, Applied Geochemistry, vol. 24, 3, 463-474.

Franz, H. B.; McAdam, A. C.; Ming, D. W.; Freissinet, C.; Mahaffy, P. R.; Eldridge, D. L.; Fischer, W. W.; Grotzinger, J. P.; House, C. H.; Hurowitz, J. A.; McLennan, S. M.; Schwenzer, S. P.; Vaniman, D. T.; Archer, P. D., Jr.; Atreya, S. K.; Conrad, P. G.; Dottin, J. W., III; Eigenbrode, J. L.; Farley, K. A.; Glavin, D. P.; Johnson, S. S.; Knudson, C. A.; Morris, R. V.; Navarro-González, R.; Pavlov, A. A.; Plummer, R.; Rampe, E. B.; Stern, J. C.; Steele, A.; Summons, R. E.; Sutter, B., 2017, Large sulfur isotope fractionations in Martian sediments at Gale crater, Nature Geoscience, Volume 10, Issue 9, pp. 658-662.

Gendrin, Aline; Mangold, Nicolas; Bibring, Jean-Pierre; Langevin, Yves; Gondet, Brigitte; Poulet, François; Bonello, Guillaume; Quantin, Cathy; Mustard, John; Arvidson, Ray; Le Mouélic, Stéphane, Sulfates in Martian Layered Terrains: The OMEGA/Mars Express View, Science, Volume 307, Issue 5715, pp. 1587-1591.

Glotch, Timothy D.; Rogers, A. Deanne, 2013, Evidence for magma-carbonate interaction beneath Syrtis Major, Mars, Journal of Geophysical Research: Planets, Volume 118, Issue 1, pp. 126-137.

Griffith, L. L.; Shock, E. L., 1995, A geochemical model for the formation of hydrothermal carbonates on Mars, Nature, 377, 6548, 406-408.

Grott, M.; Morschhauser, A.; Breuer, D.; Hauber, E., 2011, Volcanic outgassing of $CO_2$ and $H_2O$ on Mars, Earth Planet. Sci. Lett., Volume 308, Issue 3, p. 391-400.

Grotzinger, J. P.; Arvidson, R. E.; Bell, J. F.; Calvin, W.; Clark, B. C.; Fike, D. A.; Golombek, M.; Greeley, R.; Haldemann, A.; Herkenhoff, K. E.; Jolliff, B. L.; Knoll, A. H.; Malin, M.; McLennan, S. M.; Parker, T.; Soderblom, L.; Sohl-Dickstein, J. N.; Squyres, S. W.; Tosca, N. J.; Watters, W. A., 2005, Stratigraphy and sedimentology of a dry to wet eolian depositional system, Burns formation, Meridiani Planum, Mars, Earth Planet. Sci. Lett., 240, 1, 11-72.

Grotzinger, J.P., Milliken, R.E., 2012. The sedimentary rock record of Mars: Distribution, origins, and global stratigraphy. In: Grotzinger, J.P. (Ed.), Sedimentary Geology of Mars, Special Publications, vol. 102. SEPM (Society for Sedimentary Geology), 1–48.

Haberle, R. M.; Catling, D. C.; Carr, M. H.; Zahnle, K. J., 2017, The Early Mars Climate System, in The atmosphere and climate of Mars, Edited by R.M. Haberle et al. ISBN: 9781139060172. Cambridge University Press, 2017, 497-525.

Hanna, J. C.; Phillips, R. J., 2005, Hydrological modeling of the Martian crust with application to the pressurization of aquifers, Journal of Geophysical Research, 110, E1, CiteID E01004.

Harrison, K. P.; Grimm, R. E., 2009, Regionally compartmented groundwater flow on Mars, Journal of Geophysical Research, Volume 114, Issue E4, CiteID E04004.

Hurowitz, J.A. et al., 2010. Origin of acidic surface waters and the evolution of atmospheric chemistry on early Mars. Nat. Geosci. 3, 323–326.

Hurowitz, J. A.; Fischer, Woodward W., 2014, Contrasting styles of water-rock interaction at the Mars Exploration Rover landing sites, Geochim. Cosmocim. Acta,127, 25-38.

Hynek, Brian M.; Phillips, Roger J., 2008, The stratigraphy of Meridiani Planum, Mars, and implications for the layered deposits' origin, Earth Planet. Sci. Lett., 274, 214-220.

Jakosky, Bruce M.; Edwards, Christopher S., 2018, Inventory of $CO_2$ available for terraforming Mars, Nature Astronomy, 2, 634-639.

Jakosky, B. M.; Brain, D.; Chaffin, M.; Curry, S.; Deighan, J.; Grebowsky, J.; Halekas, J.; Leblanc, F.; Lillis, R.; Luhmann, J. G.; Andersson, L.; Andre, N.; Andrews, D.; Baird, D.; Baker, D.; Bell, J.; Benna, M.; Bhattacharyya, D.; Bougher, S.; Bowers, C.; Chamberlin, P.; Chaufray, J.-Y.; Clarke, J.; Collinson, G.; Combi, M.; Connerney, J.; Connour, K.; Correira, J.; Crabb, K.; Crary, F.; Cravens, T.; Crismani, M.; Delory, G.; Dewey, R.; DiBraccio, G.; Dong, C.; Dong, Y.; Dunn, P.; Egan, H.; Elrod, M.; England, S.; Eparvier, F.; Ergun, R.; Eriksson, A.; Esman, T.; Espley, J.; Evans, S.; Fallows, K.; Fang, X.; Fillingim, M.; Flynn, C.; Fogle, A.; Fowler, C.; Fox, J.; Fujimoto, M.; Garnier, P.; Girazian, Z.; Groeller, H.; Gruesbeck, J.; Hamil, O.; Hanley, K. G.; Hara, T.; Harada, Y.; Hermann, J.; Holmberg, M.; Holsclaw, G.; Houston, S.; Inui, S.; Jain, S.; Jolitz, R.; Kotova, A.; Kuroda, T.; Larson, D.; Lee, Y.; Lee, C.; Lefevre, F.; Lentz, C.; Lo, D.; Lugo, R.; Ma, Y.-J.; Mahaffy, P.; Marquette, M. L.; Matsumoto, Y.; Mayyasi, M.; Mazelle, C.; McClintock, W.; McFadden, J.; Medvedev, A.; Mendillo, M.; Meziane, K.; Milby, Z.; Mitchell, D.; Modolo, R.; Montmessin, F.; Nagy, A.; Nakagawa, H.; Narvaez, C.; Olsen, K.; Pawlowski, D.; Peterson, W.; Rahmati, A.; Roeten, K.; Romanelli, N.; Ruhunusiri, S.; Russell, C.; Sakai, S.; Schneider, N.; Seki, K.; Sharrar, R.; Shaver, S.; Siskind, D. E.; Slipski, M.; Soobiah, Y.; Steckiewicz, M.; Stevens, M. H.; Stewart, I.; Stiepen, A.; Stone, S.; Tenishev, V.; Terada, N.; Terada, K.; Thiemann, E.; Tolson, R.; Toth, G.; Trovato, J.; Vogt, M.; Weber, T.; Withers, P.; Xu, S.; Yelle, R.; Yiğit, E.; Zurek, R., 2018, Loss of the Martian atmosphere to space: Present-day loss rates determined from MAVEN observations and integrated loss through time, Icarus, Volume 315, p. 146-157.

Jakosky, B., 2019, The $CO_2$ inventory on Mars, Planetary & Space Science, 175, 52-59.

Kelemen, Peter B.; Matter, Juerg; Streit, Elisabeth E.; Rudge, John F.; Curry, William B.; Blusztajn, Jerzy, 2011, Rates and Mechanisms of Mineral Carbonation in Peridotite: Natural Processes and Recipes for Enhanced, in situ $CO_2$ Capture and Storage, Annual Review of Earth and Planetary Sciences, vol. 39, p.545-576.

Kite, Edwin S.; Hindmarsh, Richard C. A., 2007, Did ice streams shape the largest channels on Mars?, Geophys. Res. Lett., 34, 19, CiteID L19202.

Kite, E. S.; Halevy, Itay; Kahre, Melinda A.; Wolff, M. J.; Manga, M., 2013a, Seasonal melting and the formation of sedimentary rocks on Mars, with predictions for the Gale Crater mound, Icarus, 223,1, 181-210.

Kite, Edwin S.; Lewis, Kevin W.; Lamb, Michael P.; Newman, Claire E.; Richardson, Mark I., 2013b, Growth and form of the mound in Gale Crater, Mars: Slope wind enhanced erosion and transport, Geology, vol. 41, p. 543-546.

Kite, E.S., Williams, J.-P., Lucas, A., & Aharonson, O., 2014. Low palaeopressure of the Martian atmosphere estimated from the size distribution of ancient craters, Nature Geoscience, 7, 335-339.

Kite, E.S., Sneed, J., Mayer, D.P., Lewis, K.W., Michaels, T.I., Hore, A., & Rafkin, S.C.R., 2016. Evolution of major sedimentary mounds on Mars, J. Geophys. Res. – Planets, 121, 2282-2324, doi:10.1002/2016JE005135.

Lammer, H.; Chassefière, E.; Karatekin, Ö.; Morschhauser, A.; Niles, P. B.; Mousis, O.; Odert, P.; Möstl, U. V.; Breuer, D.; Dehant, V.; Grott, M.; Gröller, H.; Hauber, E.; Pham, Lê Binh San, 2013, Outgassing History and Escape of the Martian Atmosphere and Water Inventory, Space Science Reviews, 174, 113-154.

Malin, M.C., Edgett, K.S., 2000. Sedimentary rocks of early Mars. Science 290, 1927– 1937.

McCollom, T. M. (2018). Geochemical trends in the Burns formation layered sulfate deposits at Meridiani Planum, Mars, and implications for their origin. J. Geophys. Res.: Planets, 123, 2393–2429. https://doi.org/10.1029/2018JE005718.

McLennan, S.M. et al., 2005. Provenance and diagenesis of the evaporite-bearing Burns formation, Meridiani Planum, Mars. Earth Planet. Sci. Lett. 240, 95–121.

McLennan, S.M., Grotzinger, J.P., 2008. The sedimentary rock cycle of Mars. In: Bell, J., III (Ed.), The Martian Surface – Composition, Mineralogy, and Physical Properties. Cambridge University Press, p. 541–577.

McSween, H. Y.; Wyatt, M. B.; Gellert, R.; Bell, J. F.; Morris, R. V.; Herkenhoff, K. E.; Crumpler, L. S.; Milam, K. A.; Stockstill, K. R.; Tornabene, L. L.; Arvidson, R. E.; Bartlett, P.; Blaney, D.; Cabrol, N. A.; Christensen, P. R.; Clark, B. C.; Crisp, J. A.; Des Marais, D. J.; Economou, T.; Farmer, J. D.; Farrand, W.; Ghosh, A.; Golombek, M.; Gorevan, S.; Greeley, R.; Hamilton, V. E.; Johnson, J. R.; Joliff, B. L.; Klingelhöfer, G.; Knudson, A. T.; McLennan, S.; Ming, D.; Moersch, J. E.; Rieder, R.; Ruff, S. W.; Schröder, C.; de Souza, P. A.; Squyres, S. W.; Wänke, H.; Wang, A.; Yen, A.; Zipfel, J., 2006, Characterization and petrologic interpretation of olivine-rich basalts at Gusev Crater, Mars, J. Geophys. Res., 111, E2, CiteID E02S10.

Melwani Daswani, M.; Schwenzer, S. P.; Reed, M. H.; Wright, I. P.; Grady, M. M., 2016, Alteration minerals, fluids, and gases on early Mars: Predictions from 1-D flow geochemical modeling of mineral assemblages in meteorite ALH 84001, Meteoritics & Planetary Science, 51, 2154-2174.

Michalski, J.; Niles, P. B., 2012, Atmospheric origin of Martian interior layered deposits: Links to climate change and the global sulfur cycle, Geology, vol. 40, 5, 419-422

Minissale, A.; Vaselli, O.; Chandrasekharam, D.; Magro, G.; Tassi, F.; Casiglia, A., 2000, Origin and evolution of 'intracratonic' thermal fluids from central-western peninsular India, Earth Planet. Sci. Lett., 181, 3, 377-394.

Morrison, Shaunna M.; Downs, Robert T.; Blake, David F.; Vaniman, David T.; Ming, Douglas W.; Hazen, Robert M.; Treiman, Allan H.; Achilles, Cherie N.; Yen, Albert S.; Morris, Richard V.; Rampe, Elizabeth B.; Bristow, Thomas F.; Chipera, Steve J.; Sarrazin, Philippe C.; Gellert, Ralf; Fendrich, Kim V.; Morookian, John Michael; Farmer, Jack D.; Des Marais, David J.; Craig, Patricia I., 2018, Crystal chemistry of martian minerals from Bradbury Landing through Naukluft Plateau, Gale crater, Mars, American Mineralogist, vol. 103, issue 6, pp. 857-871.

Murchie, Scott; Roach, Leah; Seelos, Frank; Milliken, Ralph; Mustard, John; Arvidson, Raymond; Wiseman, Sandra; Lichtenberg, Kimberly; Andrews-Hanna, Jeffrey; Bishop, Janice; Bibring, Jean-Pierre;

Parente, Mario; Morris, Richard, 2009, Evidence for the origin of layered deposits in Candor Chasma, Mars, from mineral composition and hydrologic modeling, Journal of Geophysical Research, Volume 114, Issue E12, CiteID E00D05.

Nie N.X., Dauphas N., Greenwood R.C. (2017) Iron and oxygen isotope fractionation during UV photooxidation: implications for early Earth and Mars. Earth Planet. Sci. Lett. 458, 179-191.

Niles, P.B., Michalski, J., 2009. Meridiani Planum sediments on Mars formed through weathering in massive ice deposits. Nat. Geosci. 2, 215–220.

Niles, Paul B.; Catling, David C.; Berger, Gilles; Chassefière, Eric; Ehlmann, Bethany L.; Michalski, Joseph R.; Morris, Richard; Ruff, Steven W.; Sutter, Brad, 2013, Geochemistry of Carbonates on Mars: Implications for Climate History and Nature of Aqueous Environments, Space Science Reviews, 174, 1-4, 301-328.

Niles, Paul B.; Michalski, Joseph; Ming, Douglas W.; Golden, D. C., 2017, Elevated olivine weathering rates and sulfate formation at cryogenic temperatures on Mars, Nature Communications, 8, id. 998.

Ogawa, Y., & Manga, M., 2007, Thermal demagnetization of Martian upper crust by magma intrusion, Geophysical Research Letters, Volume 34, Issue 16, CiteID L16302.

Okubo, Chris H.; McEwen, Alfred S., 2007 , Fracture-Controlled Paleo-Fluid Flow in Candor Chasma, Mars, Science, 315, 5814, 983-985.

Onstott, T. C.; Ehlmann, B. L.; Sapers, H.; Coleman, M.; Ivarsson, M.; Marlow, J. J.; Neubeck, A.; Niles, P., 2019, Paleo-Rock-Hosted Life on Earth and the Search on Mars: a Review and Strategy for Exploration, Astrobiology, Online Ahead of Print, https://doi.org/10.1089/ast.2018.1960

Phillips-Lander, C.M., A. S. Elwood Madden, E. M.Hausrath, M. Elwood Madden, Aqueous alteration of pyroxene in sulfate, chloride, and perchlorate brines: Implications for post-Noachian aqueous alteration on Mars, Geochimica et Cosmochimica Acta, in press, doi.org/10.1016/j.gca.2019.05.006.

Reed, M. H., 1998), Calculation of simultaneous chemical equilibria in aqueous-mineral-gas systems and its application to modeling hydrothermal processes, in Techniques in Hydrothermal Ore Deposits Geology, Reviews in Economic Geology, vol. 10, edited by J. P. Richards and P. B. Larson, 109–124, Society of Economic Geologists, Littleton, Colo.

Rubin, David M.; Fairén, A. G.; Martínez-Frías, J.; Frydenvang, J.; Gasnault, O.; Gelfenbaum, G.; Goetz, W.; Grotzinger, J. P.; Le Mouélic, S.; Mangold, N.; Newsom, H.; Oehler, D. Z.; Rapin, W.; Schieber, J.; Wiens, R. C., 2017, Fluidized-sediment pipes in Gale crater, Mars, and possible Earth analogs, Geology, vol. 45, 1, 7-10.

Ruff, S. W.; Niles, P. B.; Alfano, F.; Clarke, A. B., 2014, Evidence for a Noachian-aged ephemeral lake in Gusev crater, Mars, Geology, 42, 359-362.

Schwenzer, S. P.; Abramov, O.; Allen, C. C.; Clifford, S. M.; Cockell, C. S.; Filiberto, J.; Kring, D. A.; Lasue, J.; McGovern, P. J.; Newsom, H. E.; Treiman, A. H.; Vaniman, D. T.; Wiens, R. C., 2012, Puncturing Mars: How impact craters interact with the Martian cryosphere, Earth Planet. Sci. Lett., 335, 9-17.

Sleep, N.H.; Zoback, M.D., 2007, Did Earthquakes Keep the Early Crust Habitable?, Astrobiology, 7(6), 1023-1032.

Squyres, S. W.; Knoll, A. H.; Arvidson, R. E.; Clark, B. C.; Grotzinger, J. P.; Jolliff, B. L.; McLennan, S. M.; Tosca, N.; Bell, J. F.; Calvin, W. M.; Farrand, W. H.; Glotch, T. D.; Golombek, M. P.; Herkenhoff, K. E.; Johnson, J. R.; Klingelhöfer, G.; McSween, H. Y.; Yen, A. S., 2006, Two Years at Meridiani Planum: Results from the Opportunity Rover, Science, 313, 5792, 1403-1407.

Stanley, Ben D.; Hirschmann, Marc M.; Withers, Anthony C., 2011, $CO_2$ solubility in Martian basalts and Martian atmospheric evolution, Geochim. Cosmocim. Acta, 75, 20, 5987-6003.

Stopar, Julie D.; Jeffrey Taylor, G.; Hamilton, Victoria E.; Browning, Lauren, 2006, Kinetic model of olivine dissolution and extent of aqueous alteration on mars, Geochimica et Cosmochimica Acta, Volume 70, Issue 24, p. 6136-6152.

Taylor, G.J., and 26 others, 2006, Causes of variations in K/Th on Mars: J. Geophys. Res., v. 111, E03S06, doi: 10.1029/ 2006JE002676.

Tomkinson, Tim; Lee, Martin R.; Mark, Darren F.; Smith, Caroline L., 2013, Sequestration of Martian CO2 by mineral carbonation, Nature Communications, Volume 4, id. 2662.

Tosca, Nicholas J.; McLennan, Scott M.; Dyar, M. Darby; Sklute, Elizabeth C.; Michel, F. Marc, 2008, Fe oxidation processes at Meridiani Planum and implications for secondary Fe mineralogy on Mars, J. Geophys. Res., 113, E5, CiteID E05005

Tosca, Nicholas J.; McLennan, Scott M., 2009, Experimental constraints on the evaporation of partially oxidized acid-sulfate waters at the martian surface, Geochim. Cosmocim. Acta, 73, 4, 1205-1222.

Usui, Tomohiro; Alexander, Conel M. O'D.; Wang, Jianhua; Simon, Justin I.; Jones, John H., 2015, Meteoritic evidence for a previously unrecognized hydrogen reservoir on Mars, Earth Planet. Sci. Lett., 410, 140-151.

van Berk, W.; Fu, Y.; Ilger, J.-M., 2012, Reproducing early Martian atmospheric carbon dioxide partial pressure by modeling the formation of Mg-Fe-Ca carbonate identified in the Comanche rock outcrops on Mars, Journal of Geophysical Research, 117, E10, E10008.

Wang, A., B. L. Jolliff, Y. Liu, and K. Connor, 2016, Setting constraints on the nature and origin of the two major hydrous sulfates on Mars: Monohydrated and polyhydrated sulfates, J. Geophys. Res. Planets, 121, 678–694, doi:10.1002/2015JE004889.

Walker, B.D., McCarthy, M.D., Fisher, A.T., Guilderson, T.P., 2008, Dissolved inorganic carbon isotopic composition of low-temperature axial and ridge-flank hydrothermal fluids of the Juan de Fuca Ridge Marine Chemistry, 108, 123-136.

Wordsworth, R.D., 2016, The Climate of Early Mars, Annual Rev. Earth Planet. Sci., 44, 381-408.

Wray, J. J.; Murchie, S. L.; Bishop, J. L.; Ehlmann, B. L.; Milliken, Ralph E.; Wilhelm, Mary Beth; Seelos, Kimberly D.; Chojnacki, Matthew, 2016, Orbital evidence for more widespread carbonate-bearing rocks on Mars, J. Geophys. Res.: Planets, 121, 4, 652-677.

Yen, A. S.; Ming, D. W.; Vaniman, D. T.; Gellert, R.; Blake, D. F.; Morris, R. V.; Morrison, S. M.; Bristow, T. F.; Chipera, S. J.; Edgett, K. S.; Treiman, A. H.; Clark, B. C.; Downs, R. T.; Farmer, J. D.; Grotzinger, J. P.; Rampe, E. B.; Schmidt, M. E.; Sutter, B.; Thompson, L. M.; MSL Science Team, 2017, Multiple stages of aqueous alteration along fractures in mudstone and sandstone strata in Gale Crater, Mars, Earth Planet. Sci. Lett., 471, 186-198.

Zabrusky, K., Andrews-Hanna, J.C., Wiseman, S.M., 2012. Reconstructing the distribution and depositional history of the sedimentary deposits of Arabia Terra, Mars. Icarus 220, 311–330.

Zolotov, M. Y., and M. V. Mironenko, 2016, Chemical models for Martian weathering profiles: Insights into formation of layered phyllosilicate and sulfate deposits, Icarus, 275, 203–220, doi:10.1016/j.icarus.2016.04.011.

**Supplementary Information for "Geochemistry constrains global hydrology on Early Mars".**

Edwin S. Kite[1] and Mohit Melwani Daswani[1,2]
1. University of Chicago, Chicago, IL.
2. Now at: Jet Propulsion Laboratory, Caltech, Pasadena, CA.

## 1. Detailed description of the CHIM-XPT modeling runs

We used program CHIM-XPT (Reed, 1998) to compute the equilibrium aqueous alteration of Mars basalt and the leaching of ions in a 1-D fluid-centered flow-through pathway through the basalt or olivine-only aquifer (Supplementary Figure 1). The leading fluid parcel reacts and equilibrates with unreacted rock as it moves along the flow path, and is out of contact and out of equilibrium with the preceding rock parcels, so back-reaction of the moving parcel of fluid with the previously altered rock is prevented. Precipitated phases are fractionated from the system at each step along the path (see Bethke, 2007, p. 17; and Reed, 1998, pp. 119–120 for further details). The thermodynamic database, SolthermBRGM[6], contains equilibrium constants for a large number of minerals, chemical species and gases from 0.01 to 600°C and pressures from 1 bar to 5 kbar. SolthermBRGM includes low-temperature species from the BRGM Thermoddem database (Blanc et al., 2012) among others.

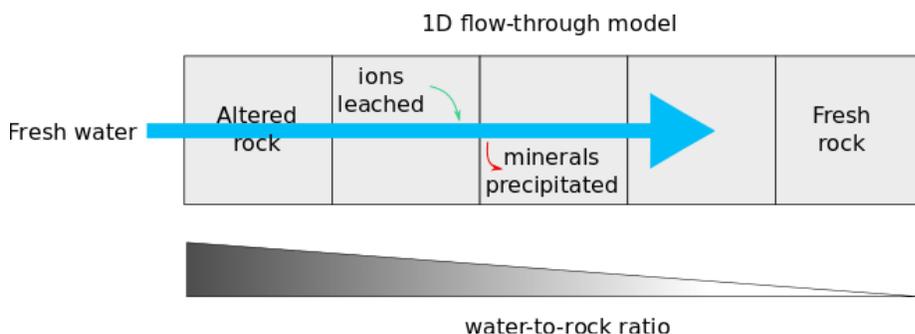

*Supplementary Figure 1. Cartoon of the reaction path model carried out with CHIM-XPT. Fresh water equilibrated with the atmosphere equilibrates with sequential parcels of unaltered rock. Along the fluid path, ions are leached from the rock into the fluid, and ions are removed from the fluid into the rock as saturated minerals are precipitated. Water-to-rock ratio decreases from left to right, as the fluid reacts with additional rock.*

The reactant rock compositions we used are described in the main text. The composition of the fluids used were initially in equilibrium with the atmosphere pressures shown in SI Table 1, which are based on linear scaling (gas volume/total volume = gas pressure/total pressure) using the current atmosphere's volume mixing ratio (Mahaffy et al., 2013), and ignoring CO because it is minor (<$10^{-3}$ volume ratio) compared to other components.

---

[6] Available at https://pages.uoregon.edu/palandri/data/solthermBRGM.xpt.

*Supplementary Table 1. Compositions of fresh fluids equilibrated with Mars atmospheres, prior to interaction with the reactant rock. $N_2$ gas exerts pressure and is soluble in the fluid, but is unreactive in our model and is not taken up by minerals.*

| Atmospheric pressure (bar) | $CO_2$ (mol/kg) | $HCO_3^-$ (mol/kg) | $CO_3^{2-}$ (mol/kg) | $N_2$ (mol/kg) | $O_2$ (mol/kg) | pH |
|---|---|---|---|---|---|---|
| $6 \times 10^{-3}$ | $3.84 \times 10^{-4}$ | $1.04 \times 10^{-5}$ | $2.45 \times 10^{-11}$ | $1.03 \times 10^{-7}$ | $1.63 \times 10^{-8}$ | 4.99 |
| $6 \times 10^{-2}$ | $3.84 \times 10^{-3}$ | $3.30 \times 10^{-5}$ | $2.48 \times 10^{-11}$ | $1.03 \times 10^{-6}$ | $1.63 \times 10^{-7}$ | 4.49 |
| $2 \times 10^{-1}$ | $1.28 \times 10^{-2}$ | $6.04 \times 10^{-5}$ | $2.51 \times 10^{-11}$ | $3.43 \times 10^{-6}$ | $5.43 \times 10^{-7}$ | 4.23 |
| $6 \times 10^{-1}$ | $3.85 \times 10^{-2}$ | $1.05 \times 10^{-4}$ | $2.54 \times 10^{-11}$ | $1.03 \times 10^{-5}$ | $1.63 \times 10^{-6}$ | 3.99 |
| 2 | $1.29 \times 10^{-1}$ | $1.93 \times 10^{-4}$ | $2.59 \times 10^{-11}$ | $3.46 \times 10^{-5}$ | $5.47 \times 10^{-6}$ | 3.73 |
| 6 | $3.91 \times 10^{-1}$ | $3.35 \times 10^{-4}$ | $2.65 \times 10^{-11}$ | $1.05 \times 10^{-4}$ | $1.65 \times 10^{-5}$ | 3.50 |

We disallowed the formation of specific minerals, aqueous species and gases, such as antigorite and $CH_4$ (Spreadsheet DISALLOWED_MINERALS.ods) because their formation are kinetically disfavored with low temperature water-rock reactions. The phases allowed to form were vetted to include only phases that would form under the pressure and temperature conditions we have modeled. A number of thermodynamically metastable phases and phases that are not abundant in natural terrestrial basalt-water systems (e.g., thaumasite, $MgHPO_4$, etc.) were allowed in the database because their formation is rapid and kinetically favored. Relevant literature about the natural and synthetic formation conditions of the phases that formed in the models is shown in Supplementary Table 2. Phases like thaumasite that have relatively limited stability fields typically recrystallize into more stable phases in time, especially if burial, heating, and complete dehydration occur. However, in the reaction path models, we quantified the carbon captured at the time that water-rock reaction occurred, and not the fate of carbon after possible recrystallization of metastable phases in time. We think this approach is adequate because the main carbon-bearing phases formed in the models are carbonates, which do not decompose and release $CO_2$ until reaching ≳450 °C (e.g. Sharp et al., 2003). While some of the phases we allowed to form in the models are unusual in basalt-water systems on Earth, purging the database in order to only include phases commonly associated with terrestrial basalt-water settings risks overlooking potential discoveries on Mars. For example, the Fe-rich amorphous materials analyzed by the *Mars Science Laboratory* throughout Gale Crater (e.g. Bish et al., 2013; Rampe et al., 2017; Vaniman et al., 2014) are clearly one or more metastable phases, that are probably largely chemically unchanged since their formation billions of years ago. The raw results of the 1-D reaction path models are shown in two supplementary spreadsheet files: one for the basalt aquifer (MARS_BASALT.ods), and another for the olivine-only (MARS_OLIVINE.ods) aquifer.

We show the pH (SI Figure 2), total dissolved aqueous components, and gas fugacities as a function of W/R ratio for each of the modeled scenarios in two animated figures compound for the alteration of the basalt (Supplementary Figure MARS_BASALT.gif) and olivine (Supplementary MARS_OLIVINE.gif) at different $CO_2$ pressures. By "total dissolved aqueous components", we refer to the total concentration of an element in solution from all the dissolved species bearing that element. For example, the $\sum$Fe curve is the sum of dissolved iron in the form of $Fe^{2+}$, $Fe(CO_3)^{2-}$, $Fe(OH)^+$, $FeSO_4(aq)$, etc. We also show the secondary minerals precipitated during alteration of the basalt (SI Figures 3 – 8) and olivine, in grams of precipitated mineral per gram of reacted fresh rock (basalt or olivine), per kilogram of water remaining in the system. In the following figures, secondary minerals precipitated in the models are grouped for clarity (e.g., phyllosilicates = kaolinite + smectite + greenalite + …). The individual minerals, their formulae and their groups are described in SI Table 2.

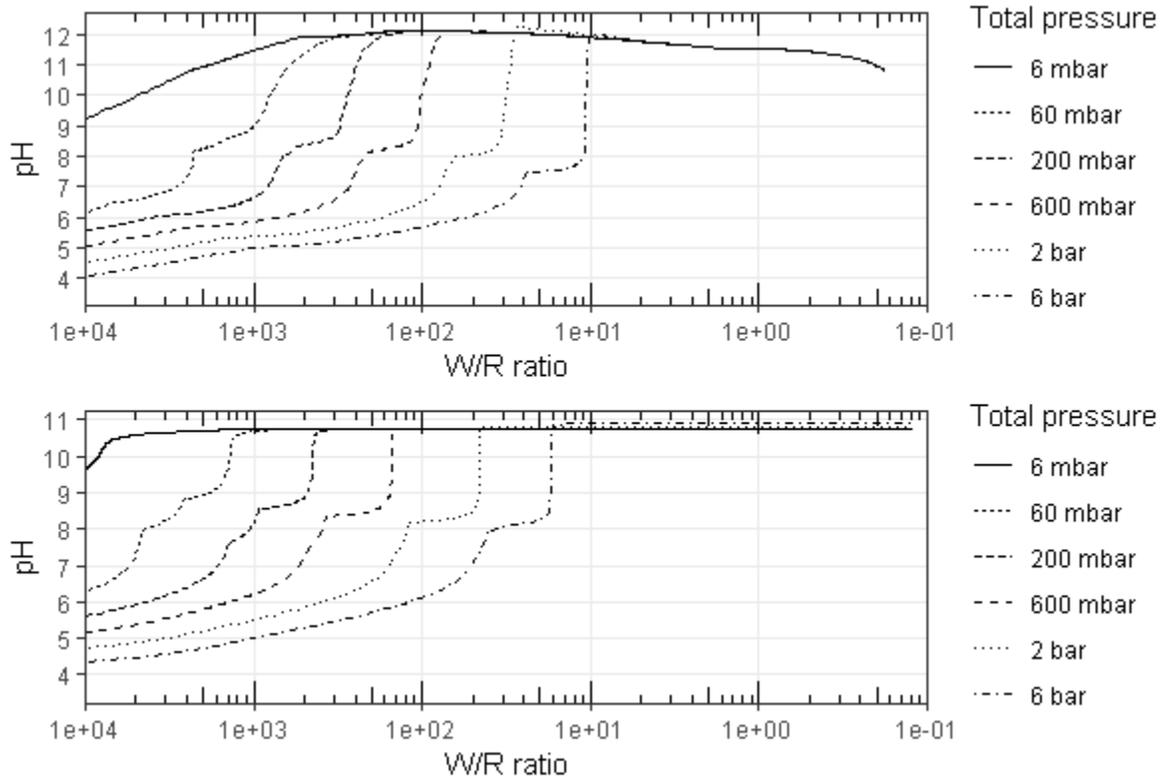

*Supplementary Figure 2. Fluid pH for all CHIM-XPT runs at different atmosphere pressures with Mars basalt (top) and Mars olivine (bottom).*

At very low W/R ratios (W/R < 1), alteration dry-out tends to occur, resulting in the secondary mineral mass exceeding the amount of fresh rock reacted. The alteration fluid at low W/R already contains a large concentration of dissolved ions, and water tends to be incorporated into minerals (phyllosilicates, zeolites, calcium-silicate-hydrate cements, and hydrous sulfates and phosphates), further increasing the concentration of ions in solution.

*Supplementary Table 2. Minerals formed in the basalt and olivine alteration models, their formulae and groups, and relevant literature about their low temperature formation conditions in natural terrestrial and laboratory settings.*

| Group | Mineral | Formula | Relevant literature about formation conditions |
|---|---|---|---|
| Phosphates | $CaAlH(PO_4)_2 \cdot 6H_2O$ | $CaAlH(PO_4)_2 \cdot 6H_2O$ | Formed within days in acidic soils (Lehr et al., 1964; Taylor et al., 1964; Taylor and Gurney, 1965, 1964) |
| | Vivianite | $Fe_3(PO_4)_2 \cdot 8H_2O$ | In lacustrine sediments (Rosenqvist, 1970), brackish/marine hypoxic waters (Dijkstra et al., 2016), with siderite in anoxic bogs (Postma, 1980) |
| | $MgHPO_4$ | $MgHPO_4$ | Most common phosphate in seawater (Atlas et al., 1976) |
| | $MnHPO_4$ | $MnHPO_4$ | Not observed naturally, but thermodynamic data lacking for Mn phosphates (serrabrancite, gatehousite, bermanite, reddingite, hureaulite). $MnHPO_4$ and related Mn phosphates are synthesized in hours to days at low T (Boonchom et al., 2008; Boonchom and Danvirutai, 2008; Evans and Sorensen, 1983) |
| | $Ca_4H(PO_4)_3{:}3H_2O$ | $Ca_4H(PO_4)_3 \cdot 3H_2O$ | Octocalcium phosphate, precursor to apatite in (e.g.) coastal and estuarine sediments (Gunnars et al., 2004; Oxmann and Schwendenmann, 2015, 2014) |
| Phyllosilicates | Kaolinite | $Al_2Si_2O_5(OH)_4$ | Common, see also chlorites below. From sedimentation of volcanic ashes in lakes, swamps, lagoons or shallow seas (Meunier, 2005, p. 312) |
| | Chamosite | $Fe_5Al(AlSi_3)O_{10}(OH)_8$ | Chlorites form slowly at low T, but precursor |

| | | |
|---|---|---|
| Clinochlore | $Mg_5Al(AlSi_3)O_{10}(OH)_8$ | phases form readily as grain coatings and pore infill, and recrystallize (Grigsby, 2001; Wilson and Pittman, 1977); authigenic in soils (Curtis et al., 1985), and in shallow marine environments (Akande and Mücke, 1993). Synthesized at low T in lab (Aja, 2002; Aja and Darby Dyar, 2002) |
| Al-free chlorite | $Mg_6Si_4O_{10}(OH)_8$ | |
| Greenalite | $Fe_3Si_2O_5(OH)_4$ | Unknown to form authigenically in soils, but gel precipitates at room temperature in anoxic water; structural reorganization and dehydration leads to crystalline greenalite (Tosca et al., 2016). |
| Minnesotaite | $Fe_3Si_4O_{10}(OH)_2$ | In lateritic weathering, in solid solution with other clay minerals (Harder, 1977). Predicted for Mars (Chevrier et al., 2007; Fairén et al., 2004) |
| K-saponite | $K_{0.33}Mg_3Al_{0.33}Si_{3.67}O_{10}(OH)_2$ | By reaction of Si+Mg-rich solutions with detrital and basaltic materials in the assemblage of salt lakes, sabkhas and alkaline lakes/swamps (Akbulut and Kadir, 2003; Meunier, 2005, pp. 307–308). Possible composition of the amorphous fraction at Yellowknife Bay on Mars (Bristow et al., 2015) |
| Na-saponite | $Na_{0.33}Mg_3Al_{0.33}Si_{3.67}O_{10}(OH)_2$ | |
| Sepiolite | $Mg_4Si_6O_{15}(OH)_2 \cdot 6H_2O$ | Precipitated in alkaline lakes/swamps (Akbulut and Kadir, 2003), and closed basin playa evaporites (Singer et al., 1998). Precipitated from seawater-like composition in lab (Baldermann et al., 2018) |
| Smectite(MX$_{80}$:5.189H$_2$O) | $Na_{0.409}K_{0.024}Ca_{0.009}(Si_{3.738}Al_{0.262})(Al_{1.598}Mg_{0.214}Fe_{0.208})O$ | From sedimentation of volcanic ashes in lakes, swamps, lagoons or shallow seas (Meunier, |

| | | | |
|---|---|---|---|
| | Mg,Na-montmorillonite | $_{10}(OH)_2 \cdot 5.189H_2O$ $Na_{0.34}Mg_{0.34}Al_{1.66}Si_4O_{10}(OH)_2$ | 2005, p. 312), in soils derived from basalt weathering (Curtin and Smillie, 1981). Synthesized in lab (Harder, 1972). Detected from orbit and in situ on Mars (Bishop et al., 2018; Clark et al., 2007) |
| | HcK-montmorillonite | $K_{0.6}Mg_{0.6}Al_{1.4}Si_4O_{10}(OH)_2$ | |
| | HcNa-montmorillonite | $Na_{0.6}Mg_{0.6}Al_{1.4}Si_4O_{10}(OH)_2$ | |
| Sulfides | Pyrite | $FeS_2$ | Most thermodynamically stable Fe-disulfide in sediments, formed by precursor amorphous FeS converting to $FeS_2$ (Schoonen, 2004). Precipitated in brackish water sediments (Postma, 1982) |
| Carbonates | Calcite Siderite Ankerite | $CaCO_3$ $FeCO_3$ $CaFe(CO_3)_2$ | Carbonate precipitation common at low temperature. Siderite precipitation may be slow, but precursor phases precipitate rapidly, and then recrystallize (Jiang and Tosca, 2019; Jimenez-Lopez and Romanek, 2004; Romanek et al., 2009) |
| Zeolites | Chabazite | $Ca(Al_2Si_4)O_{12} \cdot 6H_2O$ | Low T zeolites occurring as alteration products of volcanic and feldspathic rocks often volcanoclastics flushed by saline groundwater (Chipera and Apps, 2001; Hay and Sheppard, 2001; Ming and Boettinger, 2001; Warren, 2015, pp. 1292–1300) |
| | Ca-phillipsite | $Ca_{0.5}AlSi_3O_8 \cdot 3H_2O$ | |
| | K-phillipsite | $KAlSi_3O_8 \cdot 3H_2O$ | |
| | Na-phillipsite | $NaAlSi_3O_8 \cdot 3H_2O$ | |
| | Ca-clinoptilolite | $Ca_{0.55}(Si_{4.9}Al_{1.1})O_{12} \cdot 3.9H_2O$ | |
| Calcium-silicate-hydrates | Thaumasite | $CaSiO_3 CaSO_4 CaCO_3 \cdot 15H_2O$ | Low T seawater reaction with basalts and tuffs (Grubessi et al., 1986; Karpoff et al., 1992; Noack, 1983), and alteration of volcanic rocks in Antarctic Dry Valleys (Keys and Williams, 1981). Precipitates rapidly in lab, and destabilizes at >30 °C (Matschei et al., 2007; Pipilikaki et al., |
| | | | 2008; Schmidt et al., 2008) |
| | CSH(1.2) | $Ca_{1.2}SiO_{3.2} \cdot 2.06H_2O$ | **Tobermorite-like cement phases formed by weathering of metamorphosed carbonate + clay mineral-rich rocks (Gross, 2016, 1977). Precipitates within days in lab (Dilnesa, 2012; Gross, 1981; Schmidt et al., 2008)** |
| Sulfates | Gypsum | $CaSO_4 \cdot 2H_2O$ | **Common, especially from saturation and evaporation of saline fluid (e.g. Corselli and Aghib, 1987), but also alteration of basalt (e.g. McCanta et al., 2014)** |
| Silica | Amorphous silica | $SiO_2$ | Common. For Mars see e.g., discussion by McAdam et al. (McAdam et al., 2008). |
| Iron oxides | Goethite | FeOOH | Common |
| | Magnetite | $Fe_3O_4$ | From dissolution-precipitation of Fe-rich material and in soils (Maher and Taylor, 1988; Spiroff, 1938; Taylor et al., 1986). Precipitation from solution in lab (Hansel et al., 2005; Vayssières et al., 1998). |

## 1.1. Basalt alteration

Basalt alteration with $CO_2$-charged water causes the formation of carbonates and the drawdown of carbon under all initial $pCO_2$ conditions. Carbon drawdown begins with siderite formation and precipitation at high W/R ratios (SI Figure 3). $H^+$ produced from the dissociation of carbonic acid:

$H_2CO_3 + H_2O \rightarrow HCO_3^- + H_3O^+$ (1)

is consumed by the dissolution of olivine and pyroxene as the fluid encounters fresh rock:

$MgFeSiO_4$ (olivine) $+ 2H_3O^+ \rightarrow Mg^{2+} + Fe^{2+} H_4SiO_4 + 2OH^-$ (2)

$Mg_{0.5}Fe_{0.5}SiO_3$ (pyroxene) $+ H_3O^+ \rightarrow Mg_{0.5}^{2+} + Fe_{0.5}^{2+} + H_2SiO_3 + OH^-$ (3)

These reactions are strongly favored while the $CO_2$-charged fluid reacts with fresh rock but are limited by the amount of initial $CO_2$. When almost all carbonic acid is consumed, pH increases sharply (SI Figure 2a) as $OH^-$ produced from the breakdown of olivine and pyroxene (Reactions 2 and 3) is unbuffered by further dissociation of carbonic acid (Reaction 1). While carbonate and bicarbonate anions are supplied by the dissociation of carbonic acid, carbonates precipitate (SI Figures 3 – 8) as siderite, ankerite and calcite. At lower W/R ratios, when most carbonate is removed along the fluid path, the fluid becomes reducing, when the predominant phyllosilicates minnesotaite ($Fe_3Si_4O_{10}(OH)_2$) and $Na_{0.409}K_{0.024}Ca_{0.009}(Si_{3.738}Al_{0.262})(Al_{1.598}Mg_{0.214}Fe_{0.208})O_{10}(OH)_2 \cdot 5.189H_2O$ (a smectite of the montmorillonite group) stop precipitating in favor of greenalite ($Fe_3Si_2O_5(OH)_4$), which removes a larger amount of oxidants (namely $OH^-$) from solution per gram precipitated. Na-phillipsite ($NaAlSi_3O_8 \cdot 3H_2O$), and calcium-silicate-hydrate 1.2 ($Ca_{1.2}SiO_{3.2} \cdot 2.06H_2O$) provide the main sinks for Na, Al and Ca, while K is only weakly taken up by K-saponite ($K_{0.33}Mg_3Al_{0.33}Si_{3.67}O_{10}(OH)_2$), which allows the dissolved K concentration to increase at lower W/R ratios. Any remaining dissolved carbon is precipitated and removed as thaumasite ($Ca_3Si(OH)_6(CO_3)(SO_4) \cdot 12H_2O$). However, note that thaumasite only forms after most initial $CO_2$ is taken up by carbonates precipitated out along the reaction path (Supplementary Figures 3 – 8); carbonates require higher $pCO_2$ than thaumasite to form, and thaumasite only forms after $pCO_2$ has dropped below this level. Thaumasite is not a major $CO_2$ sink: each gram of siderite ($FeCO_3$) contains ~0.38 g of $CO_2$ whereas each gram of thaumasite ($CaSiO_3CaSO_4CaCO_3 \cdot 15H_2O$) contains ~0.07g of $CO_2$. In all the reaction path models, by far the most dramatic drop in $CO_2$ fugacity and dissolved carbon coincides with carbonate precipitation (Supplementary Figures 3 – 8).

At W/R < 10, all basalt alteration scenarios converge, yielding nearly equal pH levels, concentrations of aqueous components and precipitated minerals. Small amounts of gypsum ($CaSO_4 \cdot 2H_2O$) precipitate at W/R < 2.

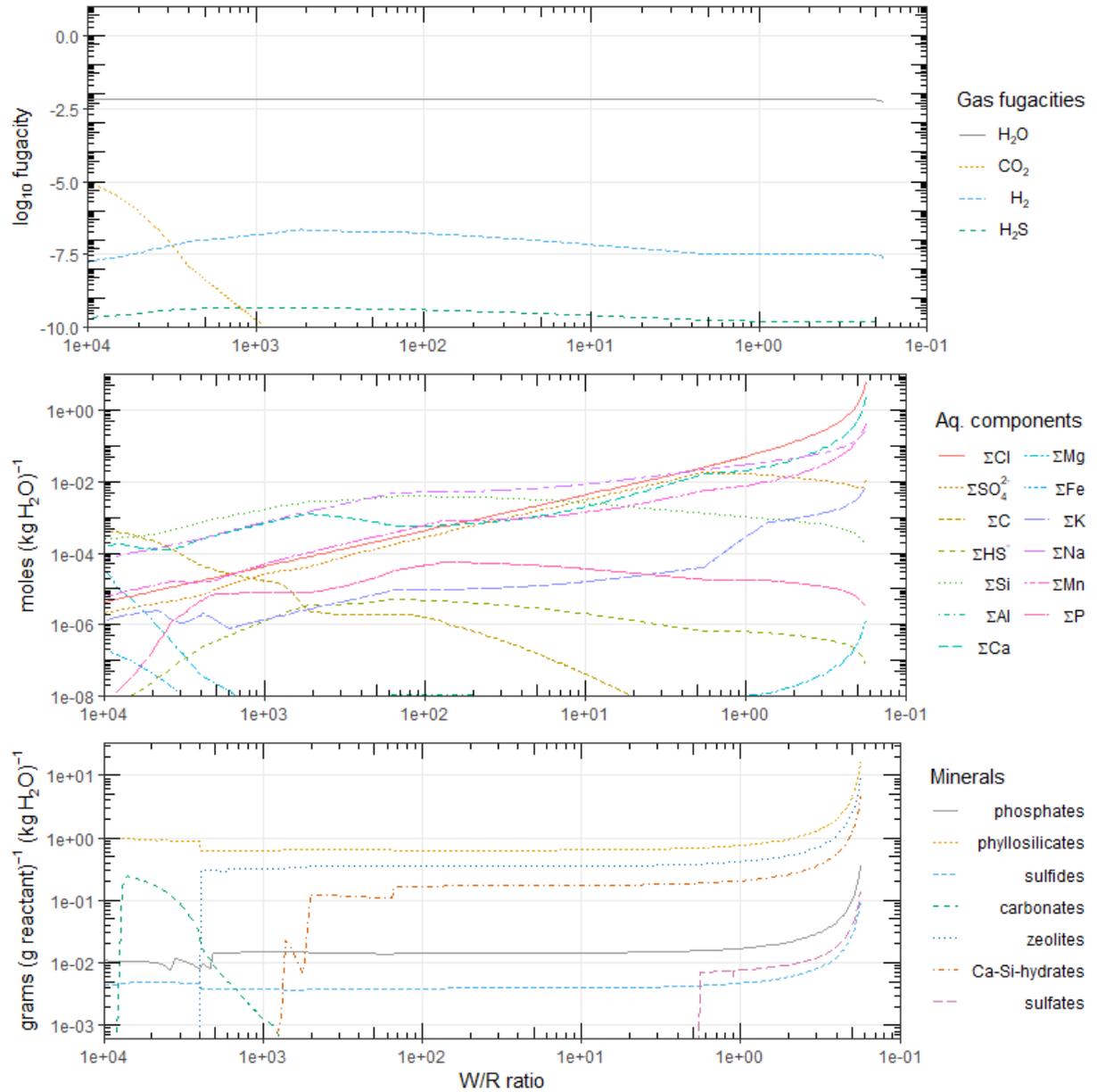

Supplementary Figure 3. Alteration of Mars basalt with fluid initially equilibrated with a 6 mbar atmosphere. Top: gas fugacities; middle: aqueous components in the fluid; bottom: secondary minerals precipitated along the alteration pathway.

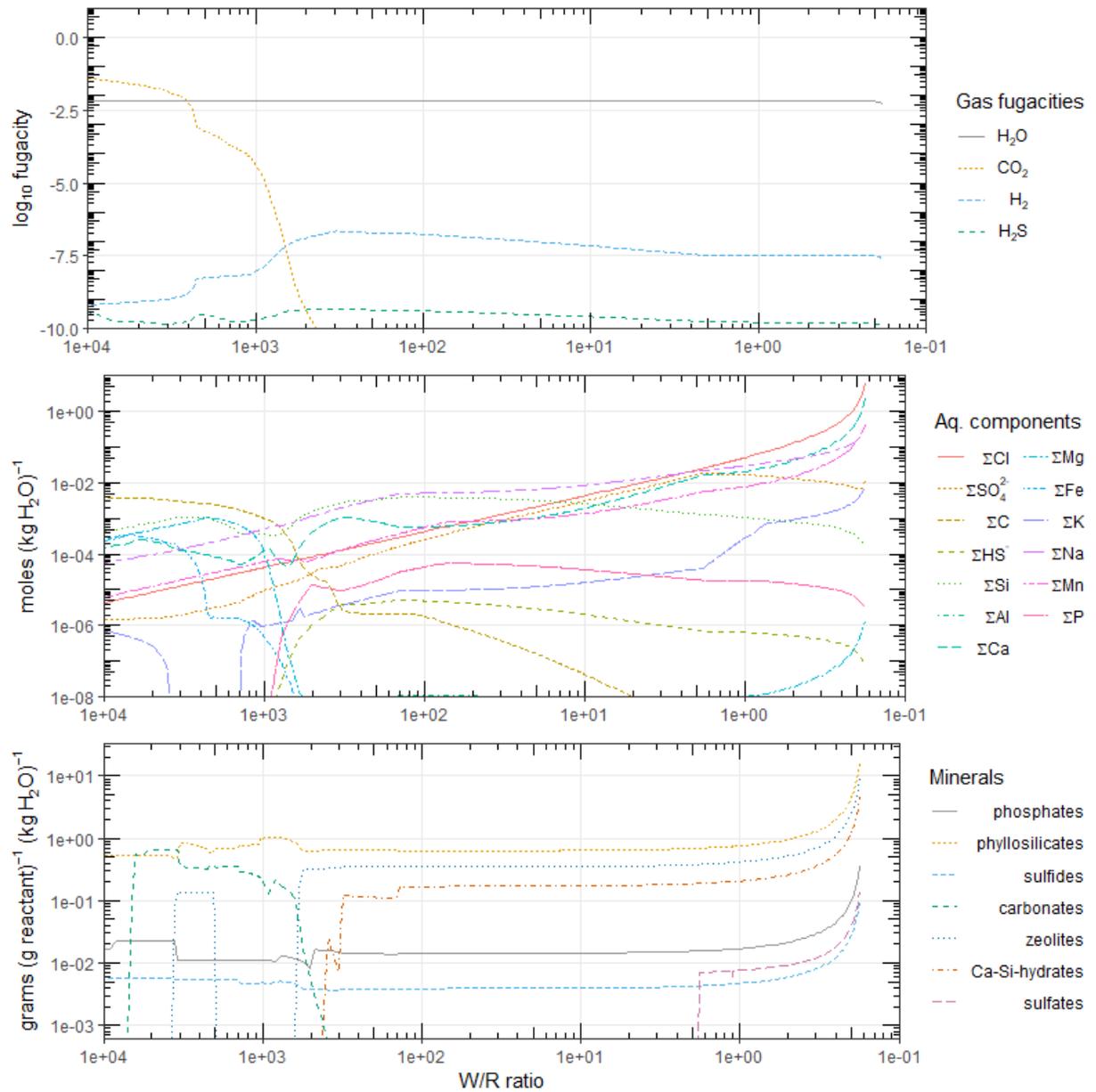

*Supplementary Figure 4. Alteration of Mars basalt with fluid initially equilibrated with a 60 mbar atmosphere. Top: gas fugacities; middle: aqueous components in the fluid; bottom: secondary minerals precipitated along the alteration pathway.*

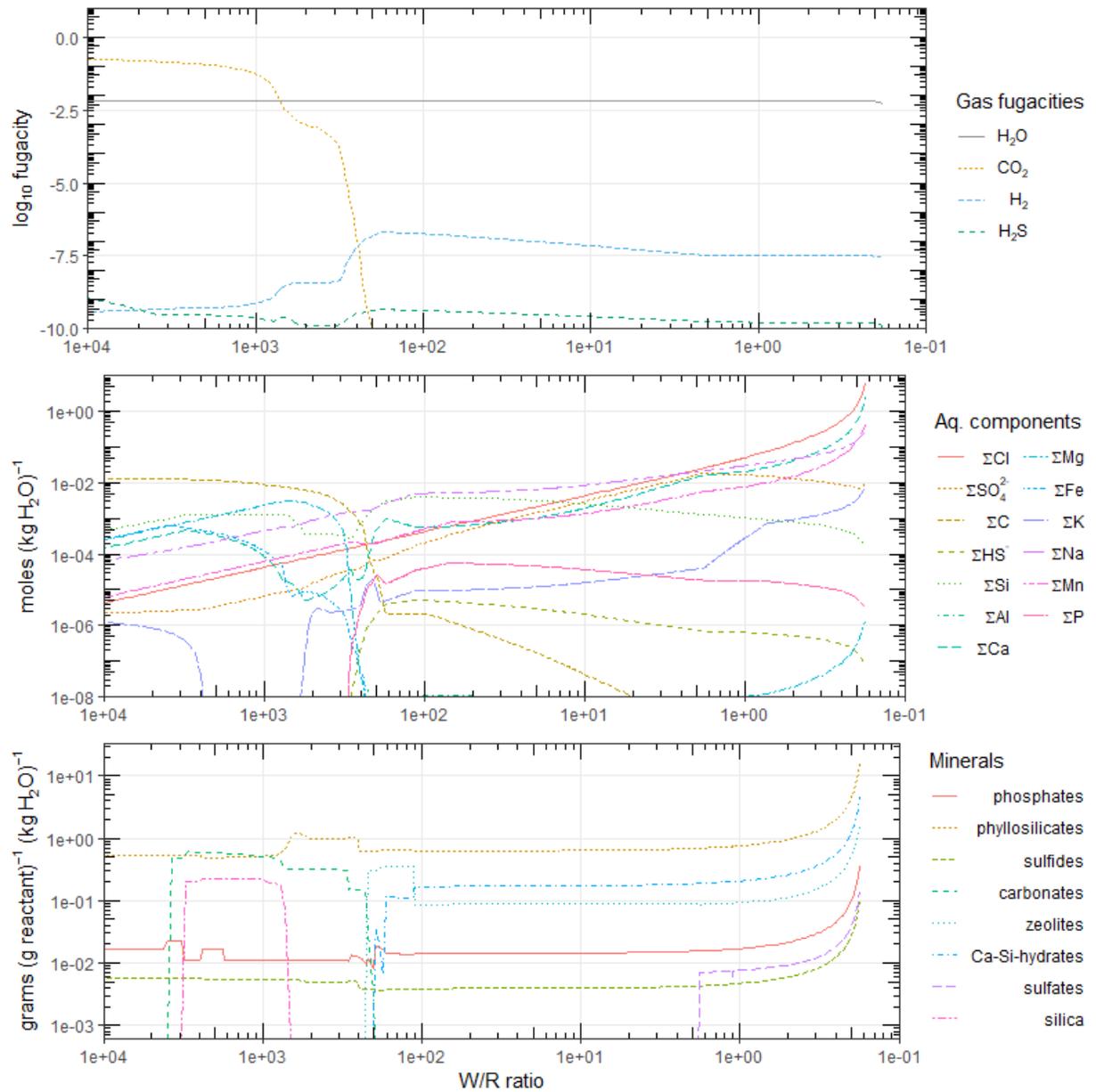

*Supplementary Figure 5. Alteration of Mars basalt with fluid initially equilibrated with a 200 mbar atmosphere. Top: gas fugacities; middle: aqueous components in the fluid; bottom: secondary minerals precipitated along the fluid pathway.*

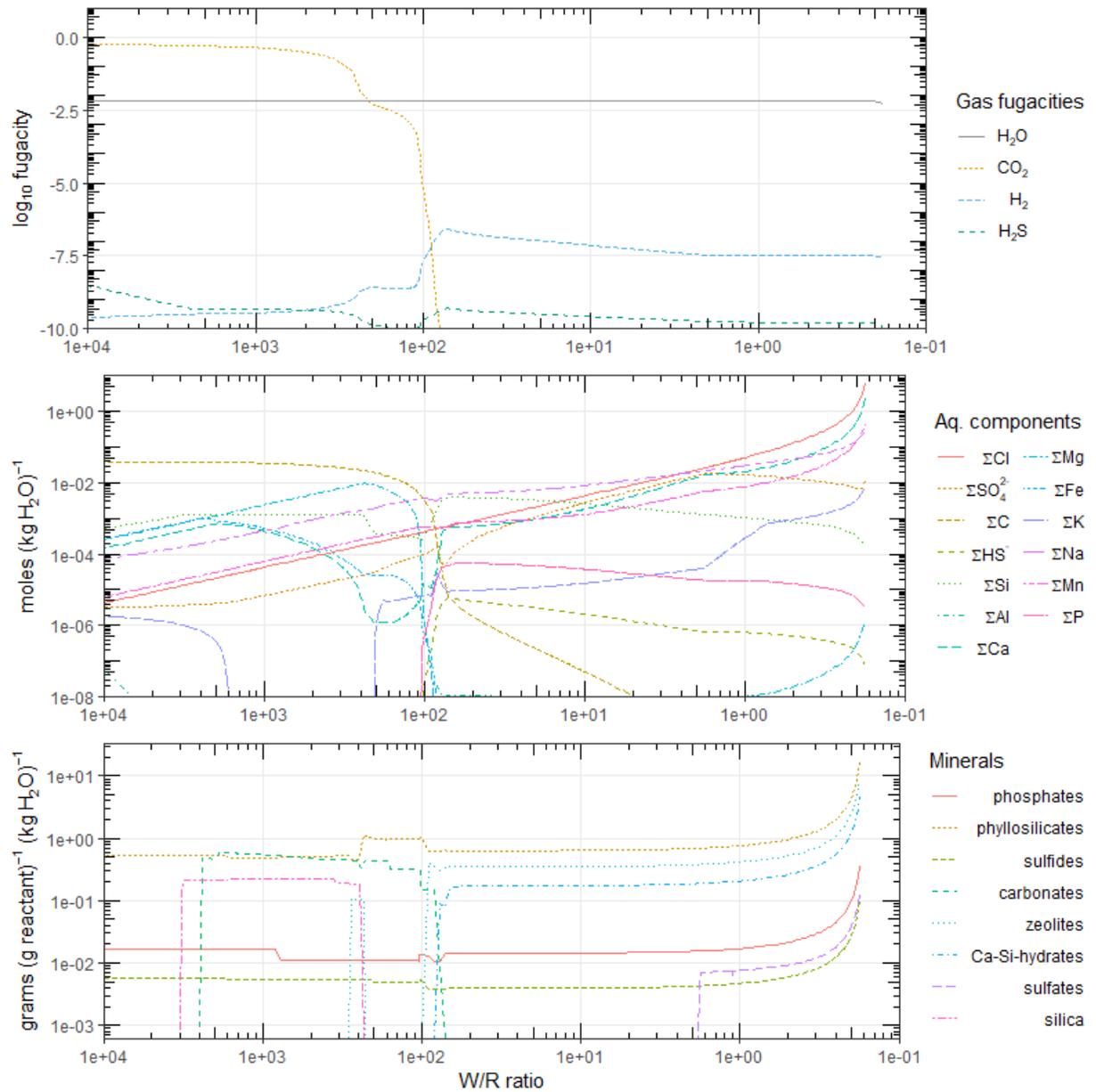

*Supplementary Figure 6. Alteration of Mars basalt with fluid initially equilibrated with a 600 mbar atmosphere. Top: gas fugacities; middle: aqueous components in the fluid; bottom: secondary minerals precipitated along the fluid pathway.*

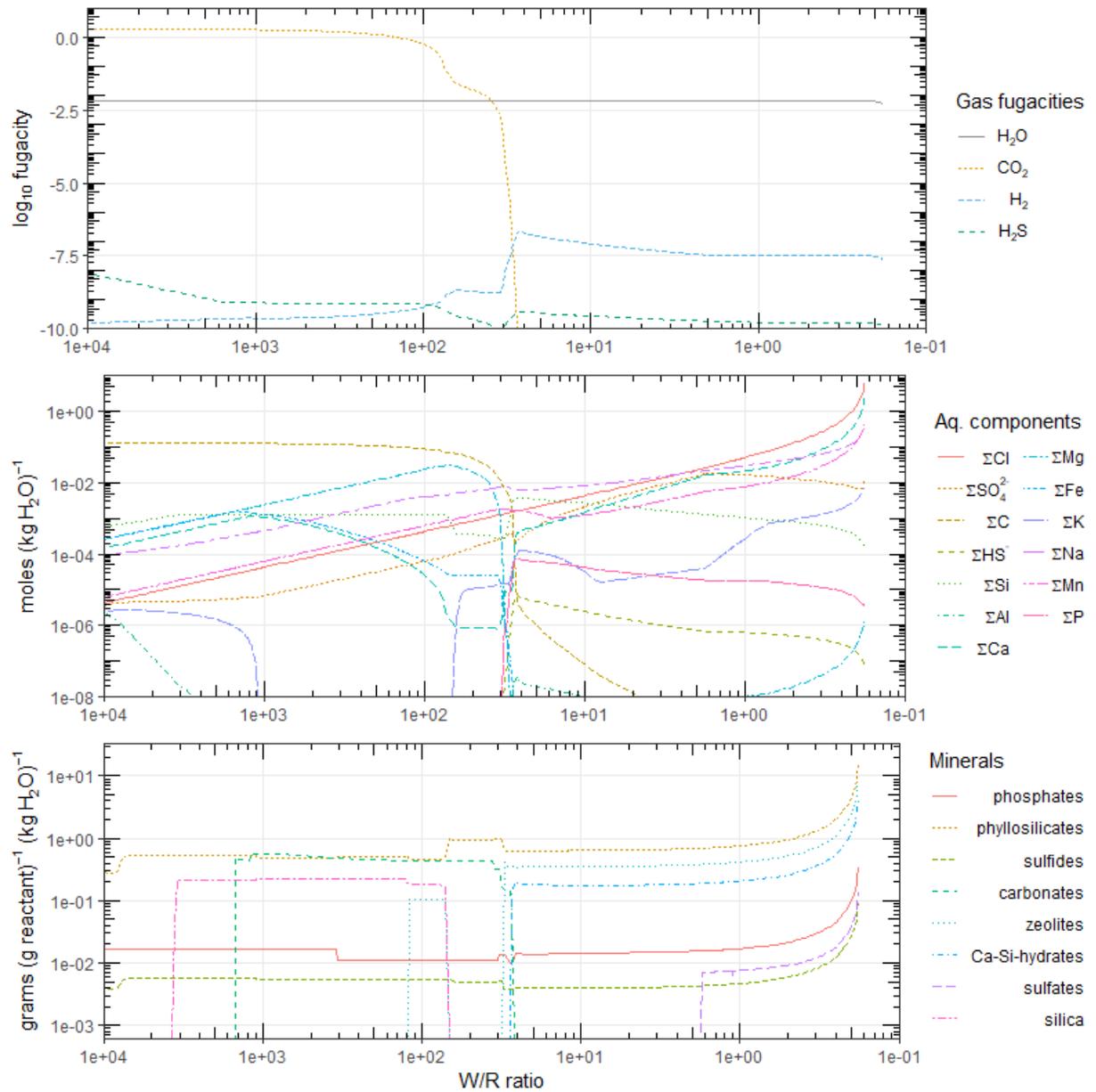

*Supplementary Figure 7. Alteration of Mars basalt with fluid initially equilibrated with a 2 bar atmosphere. Top: gas fugacities; middle: aqueous components in the fluid; bottom: secondary minerals precipitated along the fluid pathway.*

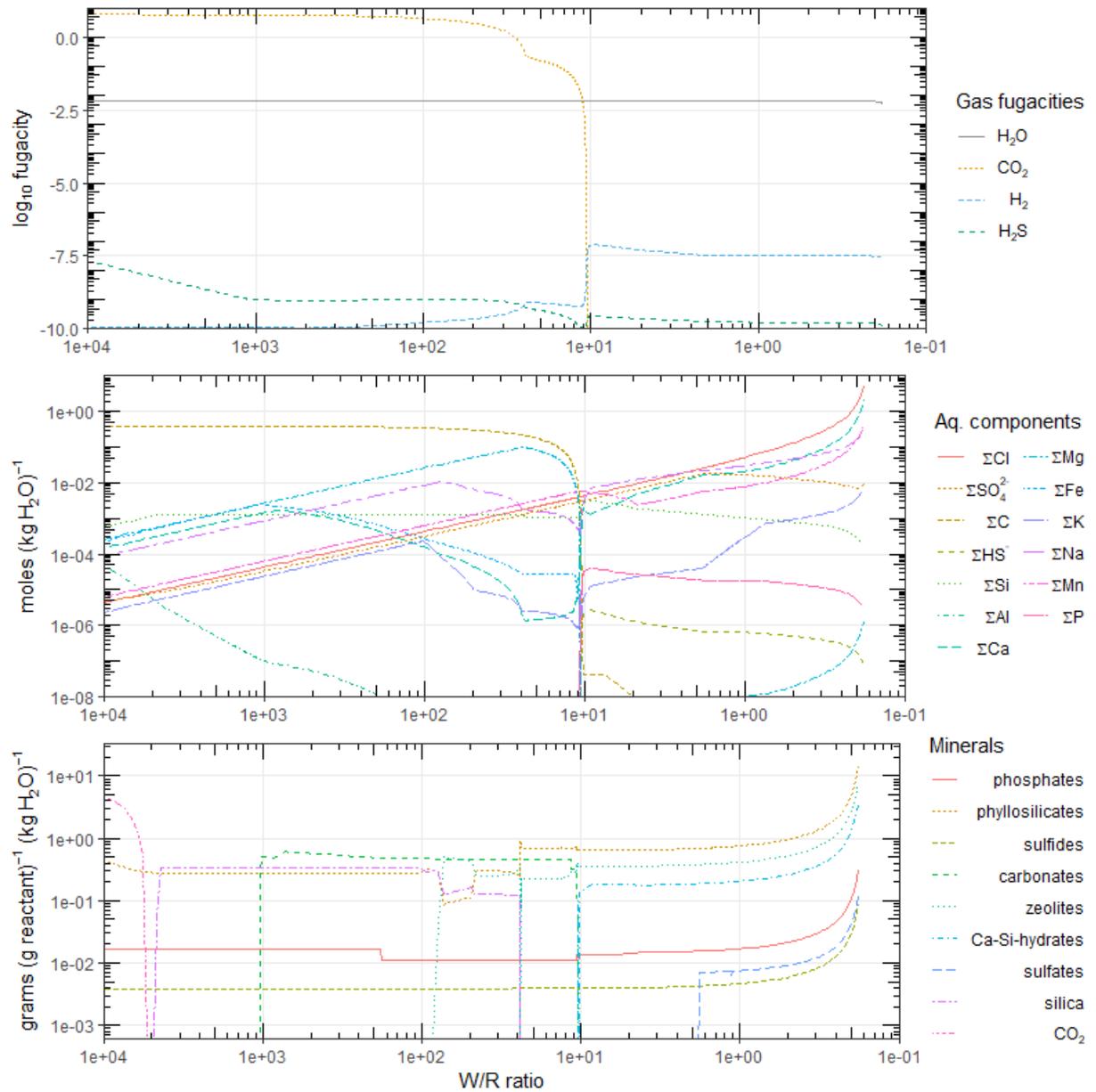

*Supplementary Figure 8. Alteration of Mars basalt with fluid initially equilibrated with a 6 bar atmosphere. Top: gas fugacities; middle: aqueous components in the fluid; bottom: secondary minerals precipitated along the fluid pathway.*

## 1.2. Olivine alteration

Reaction of olivine with the $CO_2$-bearing fluid generally follows a similar pattern to basalt alteration (SI § 1.1), in that Reaction 2 consumes acidity to produce aqueous silica and cations for carbonate minerals. Additionally, FeO dissolved from olivine drives the increase in fluid pH in a series of reactions that culminate in the production of $H_2$ and magnetite:

FeO (from olivine) + $H_2O$ → $Fe^{2+}$ + 2OH$^-$   (4)

$3Fe_2SiO_4 + 2H_2O = 3SiO_2(aq) + 2Fe_3O_4 + 2H_2(aq)$   (5)

Since the reactant olivine composition lacks calcium, the carbonates precipitated were siderite and hydromagnesite ($Mg_5(OH)_2(CO_3)_4 \cdot 4H_2O$, only at initial $pCO_2$ = 6 bar and 40 > W/R > 30). However, after most carbonate formation has occurred and pH is buffered to high levels (pH ≈ 10.8), total dissolved carbon ($\Sigma C$) plateaus at ~$5 \times 10^{-5}$ mol/kg water for all initial $pCO_2$ conditions tested; precipitation of thaumasite does not occur since it requires sulfate and calcium.

Finally, since the reactant lacks aluminum, all silica sinks from olivine alteration (amorphous silica, sepiolite, Al-free chlorite and greenalite) lack aluminum.

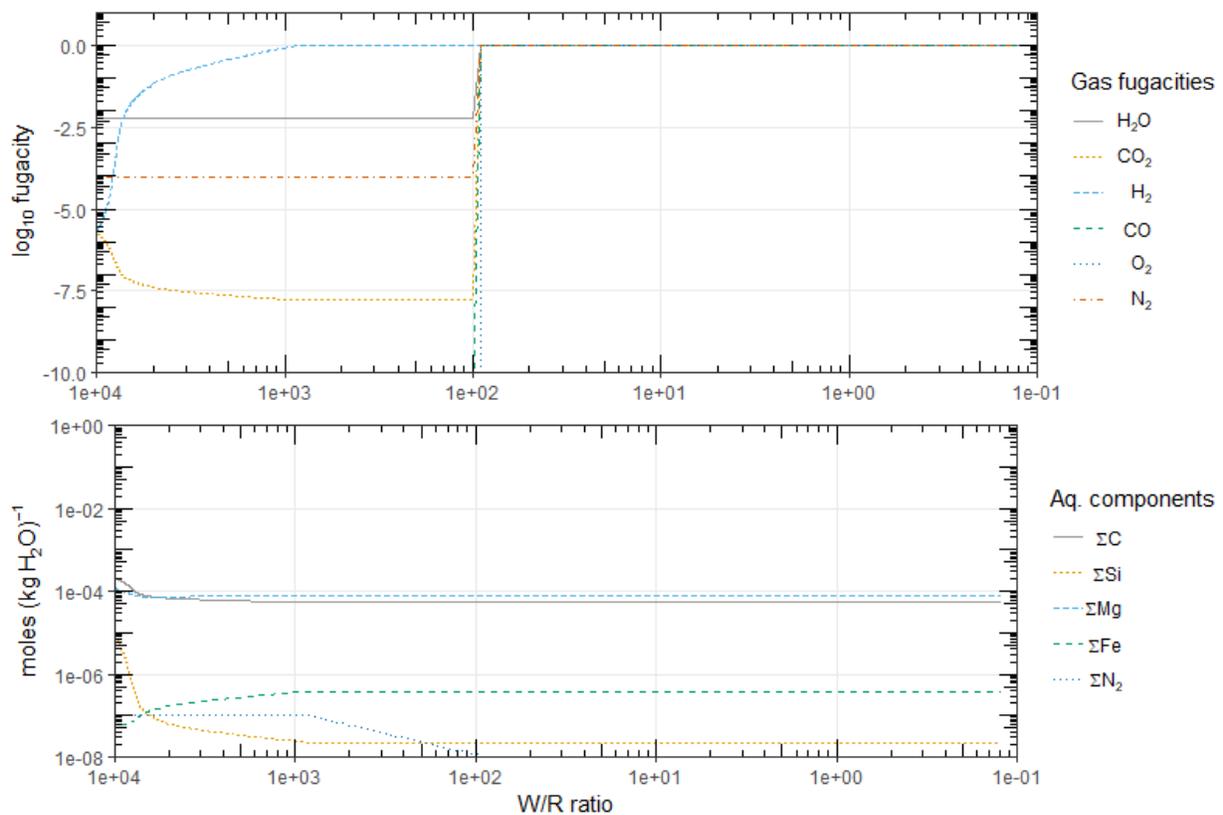

*Supplementary Figure 9. Alteration of Mars olivine with fluid initially equilibrated with a 6 mbar atmosphere. Top: gas fugacities; bottom: aqueous components in the fluid.*

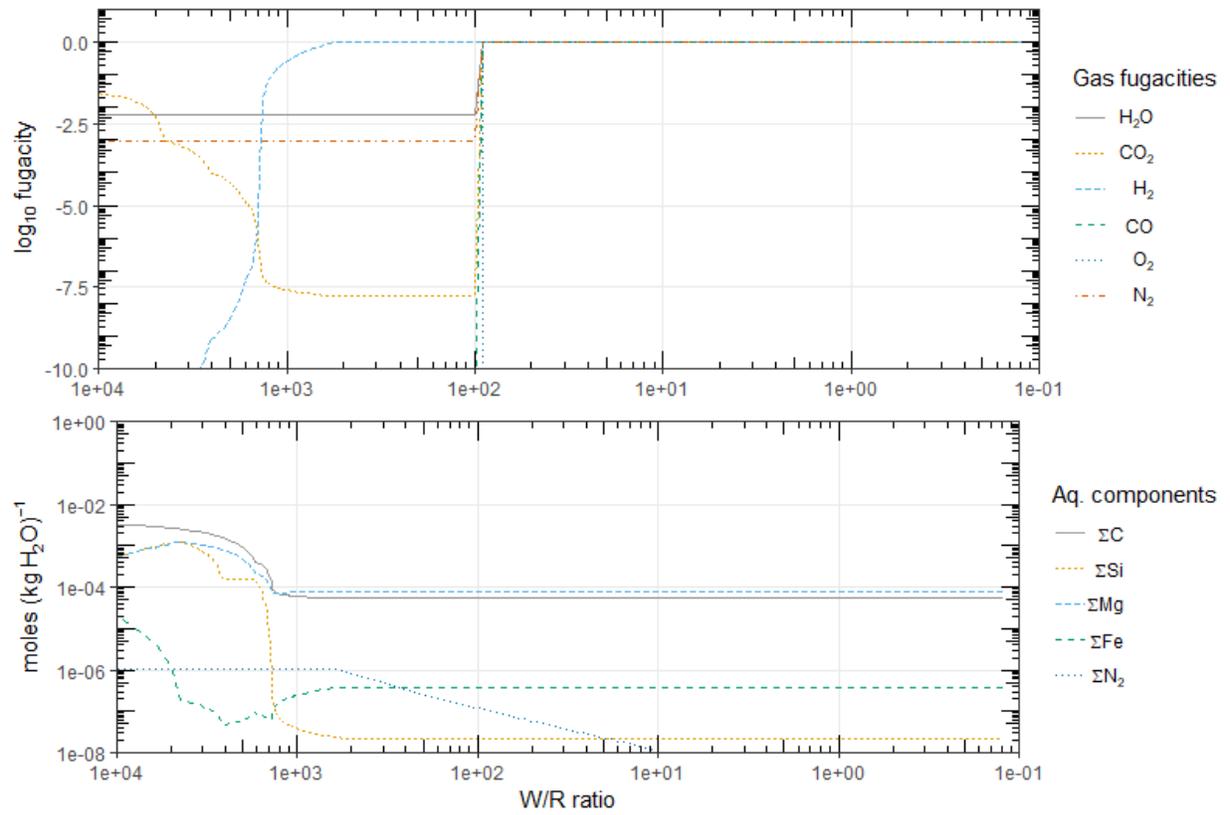

*Supplementary Figure 10. Alteration of Mars olivine with fluid initially equilibrated with a 60 mbar atmosphere. Top: gas fugacities; bottom: aqueous components in the fluid.*

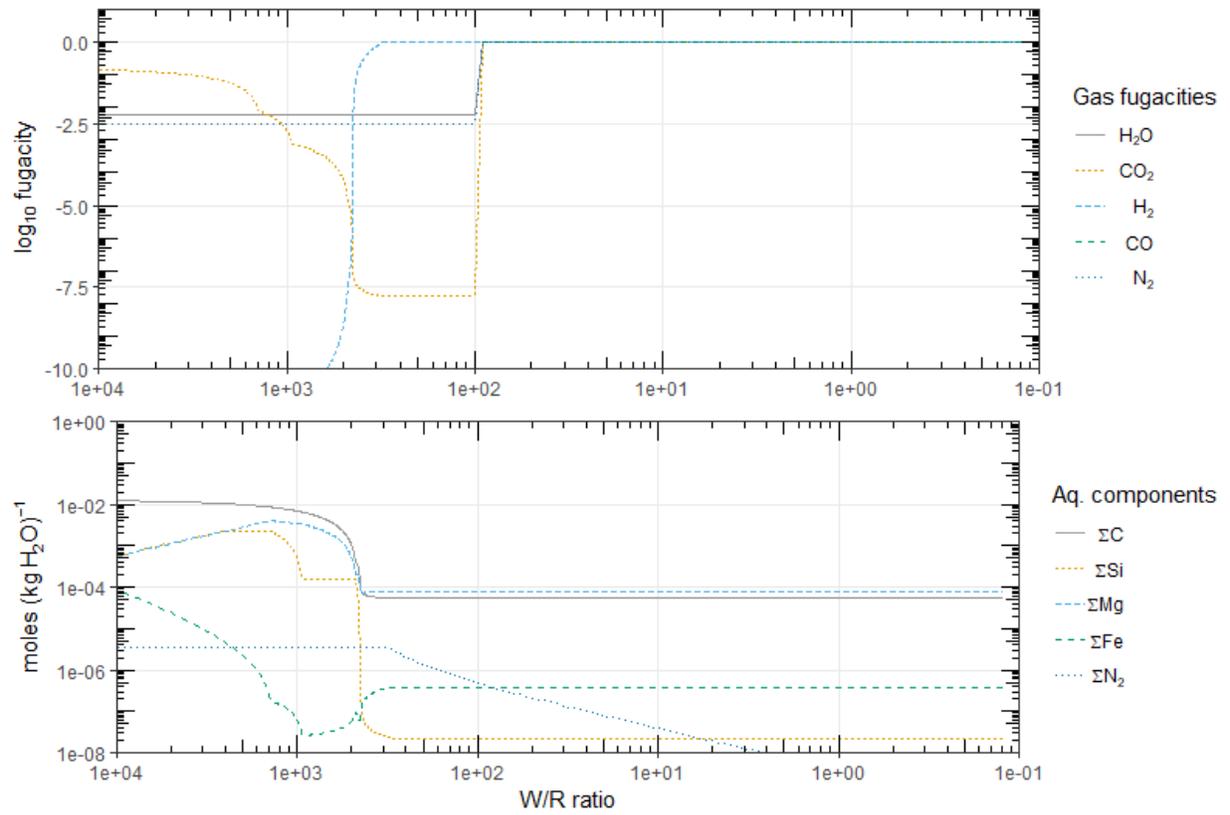

*Supplementary Figure 11. Alteration of Mars olivine with fluid initially equilibrated with a 200 mbar atmosphere. Top: gas fugacities; bottom: aqueous components in the fluid.*

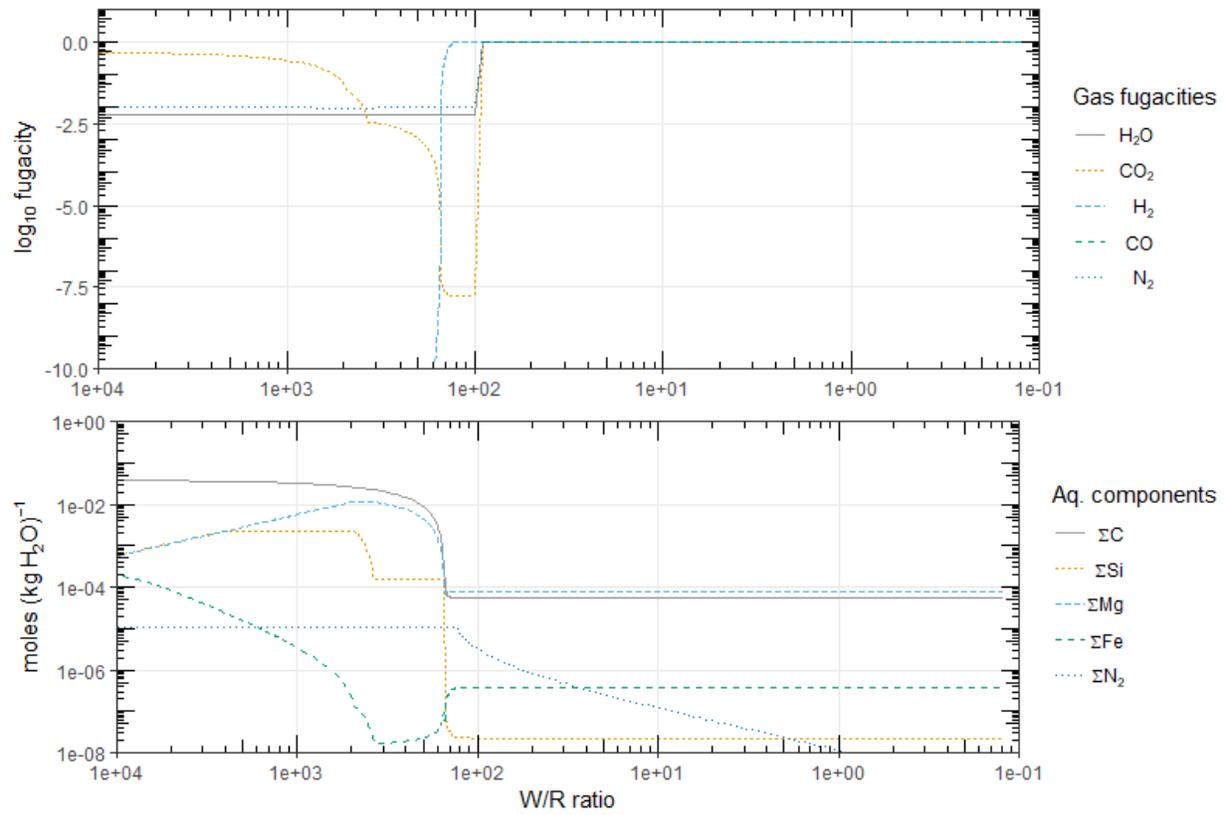

*Supplementary Figure 12. Alteration of Mars olivine with fluid initially equilibrated with a 600 mbar atmosphere. Top: gas fugacities; bottom: aqueous components in the fluid.*

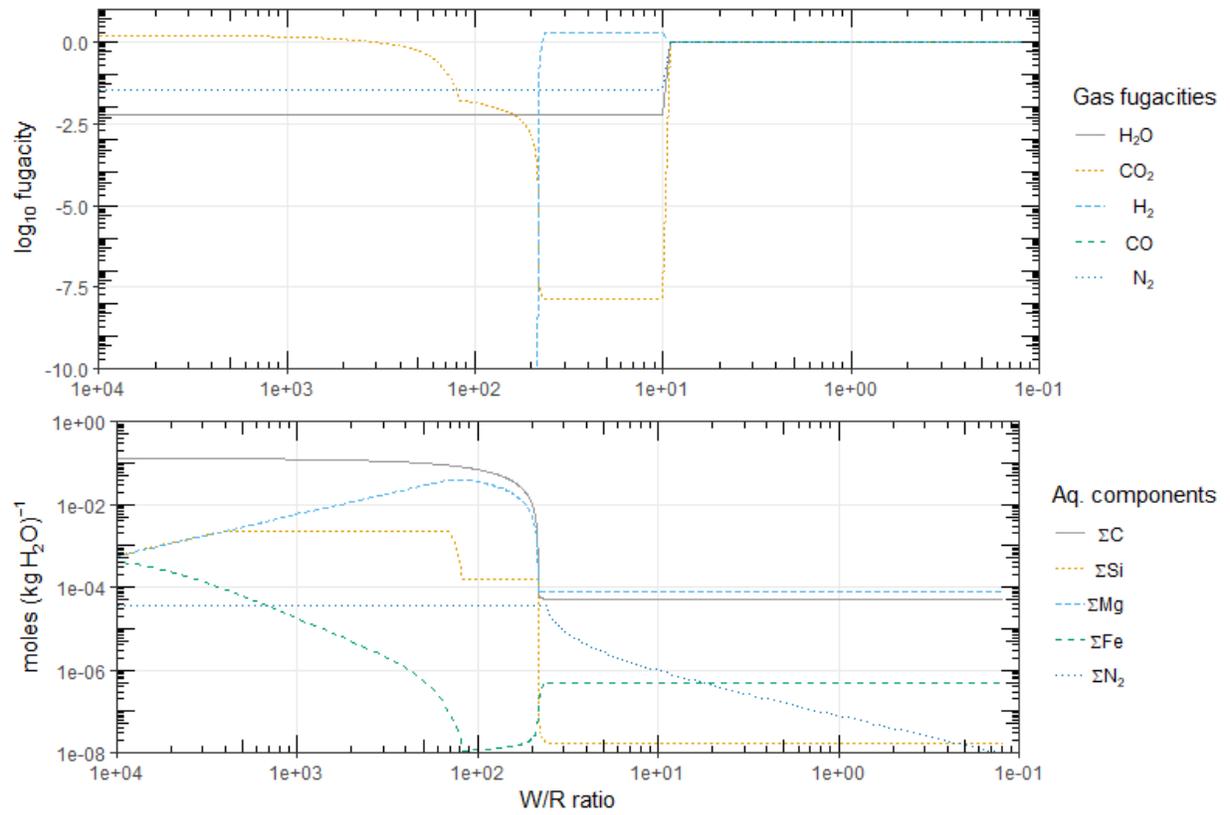

Supplementary Figure 13. Alteration of Mars olivine with fluid initially equilibrated with a 2 bar atmosphere. Top: gas fugacities; bottom: aqueous components in the fluid.

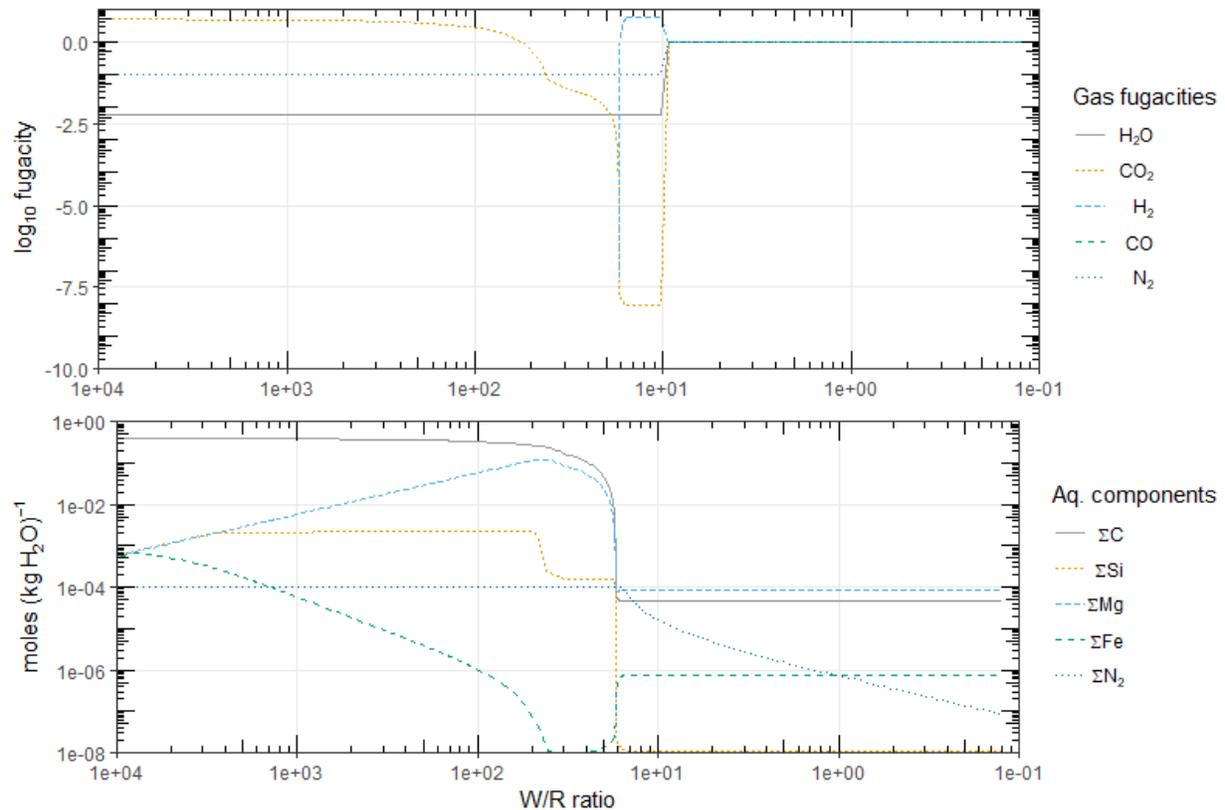

*Supplementary Figure 14. Alteration of Mars olivine with fluid initially equilibrated with a 6 bar atmosphere. Top: gas fugacities; bottom: aqueous components in the fluid.*

**References:**

Aja, S.U., 2002. The stability of Fe-Mg chlorites in hydrothermal solutions: II. Thermodynamic properties. Clays and Clay Minerals 50, 591–600.

Aja, S.U., Darby Dyar, M., 2002. The stability of Fe–Mg chlorites in hydrothermal solutions—I. Results of experimental investigations. Applied Geochemistry 17, 1219–1239. https://doi.org/10.1016/S0883-2927(01)00131-7

Akande, S.O., Mücke, A., 1993. Depositional environment and diagenesis of carbonates at the Mamu/Nkporo formation, anambra basin, Southern Nigeria. Journal of African Earth Sciences (and the Middle East) 17, 445–456. https://doi.org/10.1016/0899-5362(93)90003-9

Akbulut, A., Kadir, S., 2003. The Geology and Origin of Sepiolite, Palygorskite and Saponite in Neogene Lacustrine Sediments of the Serinhisar-Acipayam Basin, Denizli, SW Turkey. Clays and Clay Minerals 51, 279–292. https://doi.org/10.1346/CCMN.2003.0510304

Atlas, E., Culberson, C., Pytkowicz, R.M., 1976. Phosphate association with Na+, Ca2+ and Mg2+ in seawater. Marine Chemistry 4, 243–254. https://doi.org/10.1016/0304-4203(76)90011-6

Baldermann, A., Mavromatis, V., Frick, P.M., Dietzel, M., 2018. Effect of aqueous Si/Mg ratio and pH on the nucleation and growth of sepiolite at 25 °C. Geochimica et Cosmochimica Acta 227, 211–226. https://doi.org/10.1016/j.gca.2018.02.027

Bethke, C.M., 2007. Geochemical and biogeochemical reaction modeling. Cambridge University Press.

Bish, D.L., Blake, D.F., Vaniman, D.T., Chipera, S.J., Morris, R.V., Ming, D.W., Treiman, A.H., Sarrazin, P., Morrison, S.M., Downs, R.T., Achilles, C.N., Yen, A.S., Bristow, T.F., Crisp, J.A., Morookian, J.M., Farmer, J.D., Rampe, E.B., Stolper, E.M., Spanovich, N., MSL Science Team, 2013. X-ray Diffraction Results from Mars Science Laboratory: Mineralogy of Rocknest at Gale Crater. Science 341. https://doi.org/10.1126/science.1238932

Bishop, J.L., Fairén, A.G., Michalski, J.R., Gago-Duport, L., Baker, L.L., Velbel, M.A., Gross, C., Rampe, E.B., 2018. Surface clay formation during short-term warmer and wetter conditions on a largely cold ancient Mars. Nature Astronomy 2, 206–213. https://doi.org/10.1038/s41550-017-0377-9

Blanc, Ph., Lassin, A., Piantone, P., Azaroual, M., Jacquemet, N., Fabbri, A., Gaucher, E.C., 2012. Thermoddem: A geochemical database focused on low temperature water/rock interactions and waste materials. Appl. Geochem. 27, 2107–2116. https://doi.org/10.1016/j.apgeochem.2012.06.002

Boonchom, B., Danvirutai, C., 2008. A simple synthesis and thermal decomposition kinetics of MnHPO4· H2O rod-like microparticles obtained by spontaneous precipitation route. Journal of optoelectronics and advanced materials 10, 492–499.

Boonchom, B., Youngme, S., Maensiri, S., Danvirutai, C., 2008. Nanocrystalline serrabrancaite (MnPO4·H2O) prepared by a simple precipitation route at low temperature. Journal of Alloys and Compounds 454, 78–82. https://doi.org/10.1016/j.jallcom.2006.12.064

Bristow, T.F., Bish, D.L., Vaniman, D.T., Morris, R.V., Blake, D.F., Grotzinger, J.P., Rampe, E.B., Crisp, J.A., Achilles, C.N., Ming, D.W., Ehlmann, B.L., King, P.L., Bridges, J.C., Eigenbrode, J.L., Sumner, D.Y., Chipera, S.J., Moorokian, J.M., Treiman, A.H., Morrison, S.M., Downs, R.T., Farmer, J.D., Marais, D.D., Sarrazin, P., Floyd, M.M., Mischna, M.A., McAdam, A.C., 2015. The origin and implications of clay minerals from Yellowknife Bay, Gale crater, Mars†. American Mineralogist 100, 824–836. https://doi.org/10.2138/am-2015-5077CCBYNCND

Chevrier, V., Poulet, F., Bibring, J.-P., 2007. Early geochemical environment of Mars as determined from thermodynamics of phyllosilicates. Nature 448, 60–63. https://doi.org/10.1038/nature05961

Chipera, S.J., Apps, J.A., 2001. Geochemical Stability of Natural Zeolites. Reviews in Mineralogy and Geochemistry 45, 117–161. https://doi.org/10.2138/rmg.2001.45.3

Clark, B.C., Arvidson, R.E., Gellert, R., Morris, R.V., Ming, D.W., Richter, L., Ruff, S.W., Michalski, J.R., Farrand, W.H., Yen, A., Herkenhoff, K.E., Li, R., Squyres, S.W., Schröder, C., Klingelhöfer, G., Bell, J.F., 2007. Evidence for montmorillonite or its compositional equivalent in Columbia Hills, Mars. Journal of Geophysical Research: Planets 112, E06S01. https://doi.org/10.1029/2006JE002756

Corselli, C., Aghib, F.S., 1987. Brine formation and gypsum precipitation in the Bannock Basin, Eastern Mediterranean. Marine Geology 75, 185–199. https://doi.org/10.1016/0025-3227(87)90103-4

Curtin, D., Smillie, G.W., 1981. Composition and Origin of Smectite in Soils Derived from Basalt in Northern Ireland. Clays and Clay Minerals 29, 277–284. https://doi.org/10.1346/CCMN.1981.0290405

Curtis, C.D., Murchison, D.G., Berner, R.A., Shaw, H., Sarnthein, M., Durand, B., Eglinton, G., Mackenzie, A.S., Surdam, R.C., 1985. Clay Mineral Precipitation and Transformation during Burial Diagenesis [and Discussion]. Philosophical Transactions of the Royal Society of London. Series A, Mathematical and Physical Sciences 315, 91–105. https://doi.org/10.1098/rsta.1985.0031

Dijkstra, N., Slomp, C.P., Behrends, T., 2016. Vivianite is a key sink for phosphorus in sediments of the Landsort Deep, an intermittently anoxic deep basin in the Baltic Sea. Chemical Geology 438, 58–72. https://doi.org/10.1016/j.chemgeo.2016.05.025

Dilnesa, B.Z., 2012. Fe-containing Hydrates and their Fate during Cement Hydration : Thermodynamic Data and Experimental Study. EPFL, Lausanne.

Evans, A., Sorensen, R.C., 1983. Determination of the kinnetic order and heaction parameters for Cd3(PO4)2 and MnHPO4. Communications in Soil Science and Plant Analysis 14, 773–783. https://doi.org/10.1080/00103628309367407

Fairén, A.G., Fernández-Remolar, D., Dohm, J.M., Baker, V.R., Amils, R., 2004. Inhibition of carbonate synthesis in acidic oceans on early Mars. Nature 431, 423–426. https://doi.org/10.1038/nature02911

Grigsby, J.D., 2001. Origin and Growth Mechanism of Authigenic Chlorite in Sandstones of the Lower Vicksburg Formation, South Texas. Journal of Sedimentary Research 71, 27–36. https://doi.org/10.1306/060100710027

Gross, S., 2016. Petrographic atlas of the Hatrurim Formation (No. GSI/05/2016). Geological Survey of Israel, Jerusalem.

Gross, S., 1981. Simulation of natural weathering processes in the Hatrurim Formation. Current research - Geological Survey of Israel 1981, 6–8.

Gross, S., 1977. The mineralogy of Hatrurim Formation, Israel. Geol. Surv. Israel, Bull. 70, 1–80.

Grubessi, O., Mottana, A., Paris, E., 1986. Thaumasite from the Tschwinning mine, South Africa. Tschermaks mineralogische und petrographische Mitteilungen 35, 149–156. https://doi.org/10.1007/BF01082082

Gunnars, A., Blomqvist, S., Martinsson, C., 2004. Inorganic formation of apatite in brackish seawater from the Baltic Sea: an experimental approach. Marine Chemistry 91, 15–26. https://doi.org/10.1016/j.marchem.2004.01.008

Hansel, C.M., Benner, S.G., Fendorf, S., 2005. Competing Fe(II)-Induced Mineralization Pathways of Ferrihydrite. Environ. Sci. Technol. 39, 7147–7153. https://doi.org/10.1021/es050666z

Harder, H., 1977. Clay mineral formation under lateritic weathering conditions. Clay Minerals 12, 281–288. https://doi.org/10.1180/claymin.1977.012.4.01

Harder, H., 1972. The role of magnesium in the formation of smectite minerals. Chemical Geology 10, 31–39. https://doi.org/10.1016/0009-2541(72)90075-7

Hay, R.L., Sheppard, R.A., 2001. Occurrence of Zeolites in Sedimentary Rocks: An Overview. Reviews in Mineralogy and Geochemistry 45, 217–234. https://doi.org/10.2138/rmg.2001.45.6

Jiang, C.Z., Tosca, N.J., 2019. Fe(II)-carbonate precipitation kinetics and the chemistry of anoxic ferruginous seawater. Earth and Planetary Science Letters 506, 231–242. https://doi.org/10.1016/j.epsl.2018.11.010

Jimenez-Lopez, C., Romanek, C.S., 2004. Precipitation kinetics and carbon isotope partitioning of inorganic siderite at 25°C and 1 atm. Geochimica et Cosmochimica Acta 68, 557–571. https://doi.org/10.1016/S0016-7037(03)00460-5

Karpoff, A.M., France-Lanord, C., Lothe, F., Karcher, P., 1992. Miocene Tuff from Mariana Basin, Leg 129, Site 802: A First Deep-Sea Occurrence of Thaumasite, in: Proceedings of the Ocean Drilling Program, 129 Scientific Results, Proceedings of the Ocean Drilling Program. Ocean Drilling Program. https://doi.org/10.2973/odp.proc.sr.129.113.1992

Keys, J.R., Williams, K., 1981. Origin of crystalline, cold desert salts in the McMurdo region, Antarctica. Geochimica et Cosmochimica Acta 45, 2299–2309. https://doi.org/10.1016/0016-7037(81)90084-3

Lehr, J.R., Frazier, A.W., Smith, J.P., 1964. A New Calcium Aluminum Phosphate, CaAlH(PO4)2·6H2O1. Soil Science Society of America Journal 28, 38–39. https://doi.org/10.2136/sssaj1964.03615995002800010024x

Mahaffy, P.R., Webster, C.R., Atreya, S.K., Franz, H., Wong, M., Conrad, P.G., Harpold, D., Jones, J.J., Leshin, L.A., Manning, H., Owen, T., Pepin, R.O., Squyres, S., Trainer, M., MSL Science Team, 2013. Abundance and Isotopic Composition of Gases in the Martian Atmosphere from the Curiosity Rover. Science 341, 263–266. https://doi.org/10.1126/science.1237966

Maher, B.A., Taylor, R.M., 1988. Formation of ultrafine-grained magnetite in soils. Nature 336, 368–370. https://doi.org/10.1038/336368a0

Matschei, T., Lothenbach, B., Glasser, F.P., 2007. Thermodynamic properties of Portland cement hydrates in the system CaO–Al2O3–SiO2–CaSO4–CaCO3–H2O. Cement and Concrete Research 37, 1379–1410. https://doi.org/10.1016/j.cemconres.2007.06.002

McAdam, A.C., Zolotov, M.Y., Mironenko, M.V., Sharp, T.G., 2008. Formation of silica by low-temperature acid alteration of Martian rocks: Physical-chemical constraints. Journal of Geophysical Research: Planets 113. https://doi.org/10.1029/2007JE003056

McCanta, M.C., Dyar, M.D., Treiman, A.H., 2014. Alteration of Hawaiian basalts under sulfur-rich conditions: Applications to understanding surface-atmosphere interactions on Mars and Venus†. American Mineralogist 99, 291–302. https://doi.org/10.2138/am.2014.4584

Meunier, A., 2005. Clays. Springer Science & Business Media.

Ming, D.W., Boettinger, J.L., 2001. Zeolites in Soil Environments. Reviews in Mineralogy and Geochemistry 45, 323–345. https://doi.org/10.2138/rmg.2001.45.11

Noack, Y., 1983. Occurrence of thaumasite in a seawater-basalt interaction, mururoa atoll (French Polynesia, South Pacific). Mineralogical Magazine 47, 47–50. https://doi.org/10.1180/minmag.1983.047.342.08

Oxmann, J.F., Schwendenmann, L., 2015. Authigenic apatite and octacalcium phosphate formation due to adsorption–precipitation switching across estuarine salinity gradients. Biogeosciences 12, 723–738. https://doi.org/10.5194/bg-12-723-2015

Oxmann, J.F., Schwendenmann, L., 2014. Quantification of octacalcium phosphate, authigenic apatite and detrital apatite in coastal sediments using differential dissolution and standard addition. Ocean Sci. 10, 571–585. https://doi.org/10.5194/os-10-571-2014

Pipilikaki, P., Papageorgiou, D., Teas, Ch., Chaniotakis, E., Katsioti, M., 2008. The effect of temperature on thaumasite formation. Cement and Concrete Composites 30, 964–969. https://doi.org/10.1016/j.cemconcomp.2008.09.004

Postma, D., 1982. Pyrite and siderite formation in brackish and freshwater swamp sediments. American Journal of Science 282, 1151–1183. https://doi.org/10.2475/ajs.282.8.1151

Postma, D., 1980. Formation of siderite and vivianite and the pore-water composition of a Recent bog sediment in Denmark. Chemical Geology 31, 225–244. https://doi.org/10.1016/0009-2541(80)90088-1

Rampe, E.B., Ming, D.W., Blake, D.F., Bristow, T.F., Chipera, S.J., Grotzinger, J.P., Morris, R.V., Morrison, S.M., Vaniman, D.T., Yen, A.S., Achilles, C.N., Craig, P.I., Des Marais, D.J., Downs, R.T., Farmer, J.D., Fendrich, K.V., Gellert, R., Hazen, R.M., Kah, L.C., Morookian, J.M., Peretyazhko, T.S., Sarrazin, P., Treiman, A.H., Berger, J.A., Eigenbrode, J., Fairén, A.G., Forni, O., Gupta, S., Hurowitz, J.A., Lanza, N.L., Schmidt, M.E., Siebach, K., Sutter, B., Thompson, L.M., 2017. Mineralogy of an ancient lacustrine mudstone succession from the Murray formation, Gale crater, Mars. Earth and Planetary Science Letters 471, 172–185. https://doi.org/10.1016/j.epsl.2017.04.021

Reed, M.H., 1998. Calculation of simultaneous chemical equilibria in aqueous-mineral-gas systems and its application to modeling hydrothermal processes, in: Richards, J.P. (Ed.), Techniques in Hydrothermal Ore Deposits Geology, Reviews in Economic Geology. Society of Economic Geologists, Inc., Littleton, CO, pp. 109–124.

Romanek, C.S., Jiménez-López, C., Navarro, A.R., Sánchez-Román, M., Sahai, N., Coleman, M., 2009. Inorganic synthesis of Fe–Ca–Mg carbonates at low temperature. Geochimica et Cosmochimica Acta 73, 5361–5376. https://doi.org/10.1016/j.gca.2009.05.065

Rosenqvist, I.Th., 1970. Formation of vivianite in holocene clay sediments. Lithos 3, 327–334. https://doi.org/10.1016/0024-4937(70)90039-3

Schmidt, T., Lothenbach, B., Romer, M., Scrivener, K., Rentsch, D., Figi, R., 2008. A thermodynamic and experimental study of the conditions of thaumasite formation. Cement and Concrete Research 38, 337–349. https://doi.org/10.1016/j.cemconres.2007.11.003

Schoonen, M.A.A., 2004. Mechanisms of sedimentary pyrite formation, in: Special Paper 379: Sulfur Biogeochemistry - Past and Present. Geological Society of America, pp. 117–134. https://doi.org/10.1130/0-8137-2379-5.117

Sharp, Z.D., Papike, J.J., Durakiewicz, T., 2003. The effect of thermal decarbonation on stable isotope compositions of carbonates. American Mineralogist 88, 87–92. https://doi.org/10.2138/am-2003-0111

Singer, A., Stahr, K., Zarei, M., 1998. Characteristics and origin of sepiolite (Meerschaum) from Central Somalia. Clay miner. 33, 349–362. https://doi.org/10.1180/000985598545525

Spiroff, K., 1938. Magnetite crystals from meteoric solutions. Economic Geology 33, 818–828. https://doi.org/10.2113/gsecongeo.33.8.818

Taylor, A.W., Gurney, E.L., 1965. Precipitation of Phosphate by Iron Oxide and Aluminum Hydroxide from Solutions Containing Calcium and Potassium1. Soil Science Society of America Journal 29, 18–22. https://doi.org/10.2136/sssaj1965.03615995002900010008x

Taylor, A.W., Gurney, E.L., 1964. The Dissolution of Calcium Aluminum Phosphate, CaAlH(PO4)2·6H2O1. Soil Science Society of America Journal 28, 63–64. https://doi.org/10.2136/sssaj1964.03615995002800010032x

Taylor, A.W., Gurney, E.L., Moreno, E.C., 1964. Precipitation of Phosphate from Calcium Phosphate Solutions by Iron Oxide and Aluminum Hydroxide1. Soil Science Society of America Journal 28, 49–52. https://doi.org/10.2136/sssaj1964.03615995002800010028x

Taylor, R.M., Maher, B.A., Self, P.G., 1986. Magnetite in soils: I. The synthesis of single-domain and superparamagnetic magnetite. Clay Minerals 22, 411–422. https://doi.org/10.1180/claymin.1987.022.4.05

Tosca, N.J., Guggenheim, S., Pufahl, P.K., 2016. An authigenic origin for Precambrian greenalite: Implications for iron formation and the chemistry of ancient seawater. GSA Bulletin 128, 511–530. https://doi.org/10.1130/B31339.1

Vaniman, D.T., Bish, D.L., Ming, D.W., Bristow, T.F., Morris, R.V., Blake, D.F., Chipera, S.J., Morrison, S.M., Treiman, A.H., Rampe, E.B., Rice, M., Achilles, C.N., Grotzinger, J.P., McLennan, S.M., Williams, J., Bell, J.F., Newsom, H.E., Downs, R.T., Maurice, S., Sarrazin, P., Yen, A.S., Morookian, J.M., Farmer, J.D., Stack, K., Milliken, R.E., Ehlmann, B.L., Sumner, D.Y., Berger, G., Crisp, J.A., Hurowitz, J.A., Anderson, R., Des Marais, D.J., Stolper, E.M., Edgett, K.S., Gupta, S., Spanovich, N., MSL Science Team, 2014. Mineralogy of a Mudstone at Yellowknife Bay, Gale Crater, Mars. Science 343. https://doi.org/10.1126/science.1243480

Vayssières, L., Chanéac, C., Tronc, E., Jolivet, J.P., 1998. Size Tailoring of Magnetite Particles Formed by Aqueous Precipitation: An Example of Thermodynamic Stability of Nanometric Oxide Particles. Journal of Colloid and Interface Science 205, 205–212. https://doi.org/10.1006/jcis.1998.5614

Warren, J.K., 2015. Evaporites: a geological compendium. Springer Berlin Heidelberg, New York, NY.

Wilson, M.D., Pittman, E.D., 1977. Authigenic clays in sandstones: recognition and influence on reservoir properties and paleoenvironmental analysis. Journal of Sedimentary Research 47.